\documentclass[a4paper]{ar-1col}
\pdfoutput=1

\usepackage{cite}
\usepackage[centertags]{amsmath}
\usepackage{amssymb}
\usepackage{bm}
\usepackage{bbm}
\usepackage{mathrsfs}
\usepackage{graphicx}
\usepackage{subfigure}
\usepackage{url}
\usepackage[hyperindex]{hyperref}
\usepackage{color}
\usepackage[ddmmyy,24hr]{datetime}
\usepackage{bigdelim}
\usepackage{booktabs}
\usepackage{dcolumn}
\usepackage{multirow}
\usepackage{subfigure}
\usepackage{cancel}
\usepackage{stackrel}
\usepackage{paralist}
\usepackage[inline, shortlabels]{enumitem}
\usepackage{xspace}
\newcommand{\nua}[1]{\ensuremath{\rlap{\kern-2.5pt\ensuremath{\overset{\scriptscriptstyle(-)}{\phantom{\nu}}}}{\ensuremath{{\nu}_{#1}}}}}

\begin{document}

\markboth{C. Giunti and T. Lasserre}{eV-scale Sterile Neutrinos}

\title{eV-scale Sterile Neutrinos}

\author{
Carlo Giunti$^1$ and Thierry Lasserre$^2$
\affil{$^1$Istituto Nazionale di Fisica Nucleare (INFN), Sezione di Torino, Via~P.~Giuria~1, I--10125 Torino, Italy; email:~carlo.giunti@to.infn.it}
\affil{$^2$Commissariat \`a l'\'energie atomique et aux \'energies alternatives, Centre~de~Saclay, DRF/IRFU, 91191 Gif-sur-Yvette, France; email:~thierry.lasserre@cea.fr}
}

\begin{abstract}
We address the phenomenology of light sterile neutrinos,
with emphasis on short-baseline neutrino oscillations.
After a review of the observed short-baseline neutrino oscillation anomalies,
we discuss the global fit of the data
and the current appearance-disappearance tension.
We also review briefly the effects of light sterile neutrinos
in $\beta$ decay, neutrinoless double-$\beta$ decay and cosmology.
Finally, we discuss the future perspective
in the search for the effects of eV-scale sterile neutrinos.
\end{abstract}


\maketitle

\tableofcontents

\section{Introduction}
\label{sec1}

Sterile neutrinos are hypothetical neutral leptons
that mix with the ordinary active neutrinos
$\nu_{e}$,
$\nu_{\mu}$, and
$\nu_{\tau}$,
but do not engage in the standard weak interactions.
The word ``sterile'' was first used in this sense by Bruno Pontecorvo in 1967
in an article~\cite{Pontecorvo:1967fh} where he considered for the first time
oscillations of the active neutrinos into sterile neutrinos
that cannot be detected.
Among the four known fundamental interactions,
only gravitational interactions,
that are due to the geometry of space-time,
are assumed to affect sterile neutrinos.
Hence, the existence of sterile neutrinos can have observable effects
in astrophysical environments and in cosmology.
Sterile neutrinos could have non-standard interactions that,
however, are expected to be much weaker than the standard weak interactions,
and hence they are not observable with current detectors.
The only way to reveal the existence of sterile neutrinos
in terrestrial experiments is through the effects generated by their mixing with the active neutrinos:
neutrino oscillations and mass effects
($\beta$ decay, neutrinoless double-$\beta$ decay, etc.).
The search for these effects is important for our understanding of Nature,
because sterile neutrinos are non-standard particles that can open the way
to the exploration of the new physics beyond the Standard Model.

In general,
there are no theoretical limits on the number of sterile neutrinos that can exist,
on their masses and on their mixing with the active neutrinos.
These quantities must be determined experimentally.
Hence, experiments have searched for possible signatures of
sterile neutrinos at different mass scales,
from the sub-eV range to the TeV scale accessible at LHC.
There are currently two interesting indications of the possible existence of sterile neutrinos:
\begin{enumerate*} [1) ]%
\item
the short-baseline neutrino oscillations anomalies
that require the sterile neutrinos at the eV mass scale;
\item
X-ray astrophysical observations that could be due
to the radiative decay of a keV-scale sterile neutrino
that can contribute to the Dark Matter in the Universe
(see Ref.~\cite{Adhikari:2016bei}).
\end{enumerate*}
In this review we consider only the first indication,
that started in the middle 90's with the results
of the LSND experiment~\cite{Athanassopoulos:1996jb,Aguilar:2001ty}
at LAMPF
in favor of short-baseline $\bar\nu_{\mu}\to\bar\nu_{e}$ oscillations
that require the existence of sterile neutrinos at the eV mass scale.
At the same time, a similar experiment,
KARMEN~\cite{Armbruster:2002mp}
at the spallation neutron source ISIS of the Rutherford Appleton Laboratory,
did not observe any effect, but since it had a shorter baseline
(about 18 m instead of about 30 m)
its results cannot exclude the LSND signal,
because the oscillation length can be longer than that accessible
in the KARMEN experiment.
The MiniBooNE experiment\cite{Aguilar-Arevalo:2018gpe},
operating at Fermilab since about 2005,
searched for the LSND signal with controversial results.
After a progressive decrease of interest of the high-energy community
in the sterile neutrinos indicated by the LSND signal,
the interest was revamped in 2011 with the discovery of the so-called
``reactor antineutrino anomaly''
due to a deficit of the rate of reactor antineutrino detection
in several experiments at distances between about 10 and 100 m
with respect to that predicted by new theoretical calculations
of the reactor antineutrino fluxes~\cite{Mueller:2011nm,Mention:2011rk,Huber:2011wv}.
The reactor antineutrino anomaly
sparked an intense research on the eV-scale sterile neutrinos
(see the review in Ref.~\cite{Gariazzo:2015rra})
that takes into account also the LSND signal and the less statistically significant
``Gallium neutrino anomaly'' discovered in 2005-2006~\cite{Abdurashitov:2005tb,Laveder:2007zz,Giunti:2006bj}.
Several new experiments have been proposed in the following years
and some of them have been realized or are under construction.
At the moment there is no definitive experimental result
either in favor or against the eV-scale sterile neutrinos,
but it is likely that the new experiments will reach a verdict
in the near future.

In this review,
we start in Section~\ref{sec2}
with a general description of the theory of sterile neutrinos
and the theory of neutrino oscillations due to eV-scale sterile neutrinos.
In Section~\ref{sec3} we review the above-mentioned
short-baseline neutrino oscillation anomalies
and the results of other experiments that
constrain active-sterile neutrino oscillations.
Section~\ref{sec4} is devoted to the global fit of
all the short-baseline neutrino oscillation data with the
addition of eV-scale sterile neutrinos
to the standard framework of three-neutrino ($3\nu$) mixing.
In Section~\ref{sec5} we review the non-oscillation effects of
eV-scale sterile neutrinos,
with particular attention to
$\beta$ decay, neutrinoless double-$\beta$ decay, and cosmology.
Finally,
in Section~\ref{sec6} we present the future perspectives
on the search for eV-scale sterile neutrinos,
and in Section~\ref{sec7} we draw our conclusions.

\section{Theory of sterile neutrinos}
\label{sec2}
The Standard Model of electroweak interactions
is a quantum field theory based on the invariance under the group
$\text{SU}(2)_{L} \times \text{U}(1)_{Y}$
of local gauge symmetry transformations.
The subscript $L$ indicates that the $\text{SU}(2)$ transformations
operate on the left-handed chiral components on the fields.
Every four-component fermion field $\psi$ can be splitted into its left and right chiral components:
$\psi=\psi_{L}+\psi_{R}$,
with
$\psi_{L} \equiv \left( 1 - \gamma_{5} \right) \psi / 2$
and
$\psi_{R} \equiv \left( 1 + \gamma_{5} \right) \psi / 2$,
with the $4\times4$ matrix $\gamma_{5}$ defined in terms of the four
Dirac $\gamma$ matrices
($
\gamma_{5}
\equiv
i
\gamma^{0}
\gamma^{1}
\gamma^{2}
\gamma^{3}
$,
such that $(\gamma_{5})^2=1$;
we use the notations and conventions of Ref.~\cite{FNPA-ARNPS}).
A chiral gauge theory as the Standard Model of electroweak interactions
is formulated in terms of the separate left and right chiral components of the fermion fields.
In the 60's, when the Standard Model was proposed,
neutrinos were believed to be massless,
because experiments searching for the electron neutrino mass obtained
upper limits of about 200 eV,
which is much smaller than the mass of the electron (0.5 MeV),
that is the other lightest known elementary fermion.
Therefore,
the Standard Model was formulated as a gauge chiral theory
in which the neutrino fields have only the left-handed component $\nu_{\alpha L}$\footnote{
We use the Greek indices $\alpha,\beta=e,\mu,\tau$ to denote the active lepton flavors.
}.
In this way neutrinos are massless,
because Dirac mass terms
$\overline{\nu_{\alpha R}} \nu_{\beta L}$
in the Lagrangian
require the existence of the right-handed components $\nu_{\beta R}$.
Neutrinos could also have Majorana mass terms\footnote{
$\mathcal{C}$
is the unitary $4\times4$ charge-conjugation matrix,
defined by
$
\mathcal{C}
\gamma_{\mu}^{T}
\mathcal{C}^{-1}
=
- \gamma_{\mu}
$
and
$
\mathcal{C}^{T} = - \mathcal{C}
$.
}
$\nu_{\alpha L}^{T} \mathcal{C}^{\dagger} \nu_{\beta L}$,
which involve only the left-handed fields $\nu_{\alpha L}$,
but this possibility is forbidden by the
$\text{SU}(2)_{L} \times \text{U}(1)_{Y}$
symmetries of the Standard Model.
Indeed, from the gauge quantum numbers listed in Table~\ref{tab:gauge},
one can easily compute that a Majorana mass term has
$I_{3}=1$ and $Y=-2$.
Hence it is not invariant under
$\text{SU}(2)_{L} \times \text{U}(1)_{Y}$
gauge transformations and invariance cannot be restored by coupling it with the
Higgs doublet\footnote{
Also the Dirac mass terms
$\overline{\nu_{\alpha R}} \nu_{\beta L}$
are not invariant under
$\text{SU}(2)_{L} \times \text{U}(1)_{Y}$
because they have $I_{3}=1/2$ and $Y=-1$,
but invariance can be restored by generating them from the invariant products
$\overline{\nu_{\alpha R}} i \sigma_{2} \Phi^{T} L_{\beta L}$.
In the standard Higgs mechanism of mass generation
the Dirac neutrino masses are generated from this terms by the
Higgs vacuum expectation value below the electroweak scale
of about 200 GeV.
}.
Therefore,
in the Standard Model neutrinos are massless.

\begin{table}
\caption{ \label{tab:gauge}
Gauge quantum numbers of the lepton and Higgs fields in the Extended Standard Model.
The index $\alpha=e,\mu,\tau$ denotes the active lepton flavors,
with
$\ell_{e} \equiv e$,
$\ell_{\mu} \equiv \mu$,
$\ell_{\tau} \equiv \tau$.
The index $s$ enumerates the sterile right-handed neutrinos.
$I$ is the weak isospin,
$I_{3}$ is its third component,
$Y$ is the hypercharge, and
$Q$ is the electric charge.
}
\begin{center}
\begin{tabular}{@{}cc|c|c|c|c@{}}
\hline
&
&
$I$
&
$I_{3}$
&
$Y$
&
$Q=I_{3}+\frac{Y}{2}$
\\
\hline
Lepton doublets
&
$
L_{\alpha L}
=
\begin{pmatrix}
\nu_{\alpha L} \\ \ell_{\alpha L}
\end{pmatrix}
$
&
$1/2$
&
$
\begin{matrix}
1/2 \\ -1/2
\end{matrix}
$
&
$-1$
&
$
\begin{matrix}
0 \\ -1
\end{matrix}
$
\\
\hline
Charged lepton singlets
&
$\ell_{\alpha R}$
&
$0$
&
$0$
&
$-2$
&
$-1$
\\
\hline
Higgs doublet
&
$
\Phi(x)
=
\begin{pmatrix}
\phi_{+}(x) \\ \phi_{0}(x)
\end{pmatrix}
$
&
$1/2$
&
$
\begin{matrix}
1/2 \\ -1/2
\end{matrix}
$
&
$+1$
&
$
\begin{matrix}
1 \\ 0
\end{matrix}
$
\\
\hline
Sterile neutrinos
&
$\nu_{\alpha R}$
or
$\nu_{s R}$
&
$0$
&
$0$
&
$0$
&
$0$
\\
\hline
\end{tabular}
\end{center}
\end{table}

However,
when neutrino oscillations were discovered in 1998
in the Super-Kamiokande atmospheric neutrino experiment
\cite{Fukuda:1998mi},
it became clear that the Standard Model must be extended in order to generate neutrino masses.
This can be obtained by adding right-handed neutrino fields
that are singlets under the $\text{SU}(2)_{L} \times \text{U}(1)_{Y}$
gauge transformations.
Hence, they are also called ``neutral lepton singlets''
or ``sterile neutrinos'',
because they do not take part to the Standard Model weak interactions.
Moreover,
there is no known constraint on the number of right-handed neutrino fields
and, instead of considering three right-handed fields $\nu_{\alpha R}$ with $\alpha,\beta=e,\mu,\tau$,
in a general theory we must consider
$N_{s}$ right-handed fields $\nu_{s R}$ with $s=1,\ldots,N_{s}$.
The introduction of these fields,
in the so-called ``Extended Standard Model'',
is a drastic change of the theory
with respect to the Standard Model,
because the right-handed neutrino fields can have Majorana mass terms\footnote{
It is possible to avoid the Majorana mass terms of the right-handed neutrino fields
by imposing a global lepton number conservation.
However, this is an assumption that is not justified in the Standard Model,
where the global lepton number conservation is an accidental symmetry
(in the absence of the right-handed neutrino fields).
}
$\nu_{s R}^{T} \mathcal{C}^{\dagger} \nu_{s' R}$
that are invariant under $\text{SU}(2)_{L} \times \text{U}(1)_{Y}$
gauge transformations
and the corresponding masses cannot be generated by the Standard Model Higgs mechanism.
Therefore,
in general the introduction right-handed neutrino fields implies that:
\begin{enumerate}
\item
There is some physics beyond the Standard Model.
\item
Massive neutrinos are Majorana particles.
\end{enumerate}
The massive neutrino fields
$\nu_{k L}$,
with $k=1,\ldots,3+N_{s}$
are obtained from the active and sterile flavor neutrino fields
through a unitary transformation that diagonalizes the Lagrangian mass term
(see Ref.~\cite{Gariazzo:2015rra}):
\begin{align}
\null & \null
\nu_{\alpha L}
=
\sum_{k=1}^{N}
U_{\alpha k}
\nu_{k L}
\quad
(\alpha=e,\mu,\tau)
,
\label{mix1}
\\
\null & \null
(\nu_{s R})^{C}
=
\sum_{k=1}^{N}
U_{(3+s) k}
\nu_{k L}
\quad
(s=1,\ldots,N_{s})
,
\label{mix2}
\end{align}
where $N=3+N_{s}$
and
$U$ is a unitary $N \times N$ matrix.
These equations relate the flavor basis of the active and sterile neutrino fields to the
mass basis.

There is a mechanism called ``seesaw'' that produces naturally small light neutrino masses given by the relation
$ m_{\text{light}} \sim m_{\text{D}}^2 / m_{R} $
where
$m_{\text{D}}$ is the scale of the Dirac neutrino masses generated with the standard Higgs mechanism,
and
$m_{R}$ is the scale of the masses of very heavy right-handed neutrinos
(see the review in Ref.~\cite{Drewes:2015jna}).
Since $m_{\text{D}}$ is smaller than the electroweak scale (about 200 GeV),
if $m_{R}$ is very large,
say $\sim 10^{14-15} \, \text{GeV}$ as predicted by Grand Unified Theories,
the light neutrino masses are naturally smaller than about a eV.
In this scenario the right-handed neutrinos are sterile,
but decoupled from the accessible low-energy physics.
Moreover,
it can be shown that the mixing between the light neutrinos
and the heavy right-handed neutrinos is strongly suppressed.
Therefore, although these sterile neutrinos are very important for the theory\footnote{
Heavy right-handed neutrinos are also useful for the generation of the
matter-antimatter asymmetry in the Universe through the so-called
``leptogenesis'' mechanism,
which however is also very difficult, if not impossible, to prove experimentally.
},
they do not have a phenomenological impact.
However,
if there are several right-handed neutrino fields,
not all of them have to be very heavy.
Some of them could be light and belong to low-energy new physics
beyond the Standard Model,
maybe connected to the Dark Matter in the Universe.
These neutrinos can have masses at all the currently accessible energy scales,
from some TeV down to the sub-eV mass scale.
In this review we consider sterile neutrinos at the eV mass scale,
that can generate neutrino oscillations
measurable in short-baseline neutrino oscillation experiments.

Since sterile neutrinos belong to physics beyond the Standard Model,
they do not have standard weak interactions,
but
they can have non-standard gauge interactions (see the review in Ref.~\cite{Volkas:2001zb})
\footnote{
Of course,
they have also gravitational interactions that,
in general relativity, affect all particles because of the geometry of space-time.
}.
However,
these interactions must have tiny effects on the behavior of the
Standard Model particles that we know,
since otherwise they would have been detected.
Therefore,
we can consider the sterile neutrinos as practically non-interacting
(as the name ``sterile'' invented by Bruno Pontecorvo~\cite{Pontecorvo:1967fh} indicates)
and we can consider the phenomenology of neutrino interactions
as solely due to weak interactions.

In most experiments neutrinos are detected through charged-current weak interactions
generated by the Lagrangian
\begin{equation}
\mathcal{L}_{\text{CC}}
=
-
\frac{ g }{ \sqrt{2} }
\sum_{\alpha=e,\mu,\tau}
\left(
\overline{\ell_{\alpha L}} \gamma^{\rho} \nu_{\alpha L} W_{\rho}^{\dagger}
+
\overline{\nu_{\alpha L}} \gamma^{\rho} \ell_{\alpha L} W_{\rho}
\right)
,
\label{CC1}
\end{equation}
where $W_{\rho}$ is the field of the $W$ vector boson and $g$ is a coupling constant.
These interactions allow us to distinguish the neutrino flavor by detecting the corresponding charged lepton.
Taking into account the mixing in Eq.~(\ref{mix1}),
we obtain
\begin{equation}
\mathcal{L}_{\text{CC}}
=
-
\frac{ g }{ \sqrt{2} }
\sum_{\alpha=e,\mu,\tau}
\sum_{k=1}^{N}
\left(
\overline{\ell_{\alpha L}} \gamma^{\rho} U_{\alpha k} \nu_{k L} W_{\rho}^{\dagger}
+
\overline{\nu_{k L}} U^{*}_{\alpha k} \gamma^{\rho} \ell_{\alpha L} W_{\rho}
\right)
.
\label{CC2}
\end{equation}
here one can see that although there are only three flavors
all the massive neutrinos take part to charged-current weak interactions
(if their masses are kinematically allowed).
The physical mixing is determined by the $3 \times N$ rectangular submatrix
composed by the first three rows of the matrix $U$
corresponding to $\alpha=e,\mu,\tau$.
Indeed, the mixing of the sterile fields in Eq.~(\ref{mix2}),
which is determined by the complementary rectangular submatrix,
is not observable.
Hence, from now on we will consider only the $3 \times N$ rectangular mixing matrix
in Eq.~(\ref{CC2}), keeping the same notation.

Although it is possible to work without a parameterization of the mixing matrix,
it is common and often useful to parameterize the mixing matrix
in terms of mixing angles and phases.
Complex phases generate CP violation that can be observed in neutrino oscillations.
In general,
the $3 \times N$ mixing matrix $U$ can be parameterized
in terms of
$3 + 3 N_{s}$ mixing angles
and
$3 + 3 N_{s}$ physical phases,
of which
$1 + 2 N_{s}$ are Dirac phases that exist for Dirac and Majorana neutrinos,
and
$N - 1$ are Majorana phases that exist only for Majorana neutrinos
and do not affect neutrino oscillations
(see Ref.~\cite{FNPA-ARNPS}).
A convenient choice is
\begin{equation}
U
=
\left[
W^{3N}
R^{2N}
W^{1N}
\cdots
W^{34}
R^{24}
W^{14}
R^{23}
W^{13}
R^{12}
\right]_{3 \times N}
\operatorname{diag}\!\left(
1, e^{i\lambda_{21}}, \ldots, e^{i\lambda_{N1}}
\right)
,
\label{mixmat}
\end{equation}
where
$W^{ab} = W^{ab}(\theta_{ab},\eta_{ab})$
is a unitary $N \times N$ matrix\footnote{
Its components are
$
\left[
W^{ab}(\vartheta_{ab},\eta_{ab})
\right]_{rs}
=
\delta_{rs}
+
\left( c_{ab} - 1 \right)
\left(
\delta_{ra} \delta_{sa}
+
\delta_{rb} \delta_{sb}
\right)
+
s_{ab}
\left(
e^{i\eta_{ab}} \delta_{ra} \delta_{sb}
-
e^{-i\eta_{ab}} \delta_{rb} \delta_{sa}
\right)
$,
where
$c_{ab}\equiv\cos\vartheta_{ab}$
and
$s_{ab}\equiv\sin\vartheta_{ab}$.
}
that performs a complex rotation in the $a$-$b$ plane
by a mixing angle $\theta_{ab}$ and a Dirac phase $\eta_{ab}$,
the orthogonal $N \times N$ matrix
$R^{ab}=W^{ab}(\theta_{ab},0)$
performs a real rotation in the $a$-$b$ plane,
and the square brackets with subscript $3 \times N$
indicate that the enclosed $N \times N$ matrix
is truncated to the first three rows.
The phases
$\lambda_{21}, \ldots \lambda_{N1}$,
collected in a diagonal matrix on the right
are the Majorana phases.
The parameterization in Eq.~(\ref{mixmat}) has two advantages:
\begin{enumerate}
\item
In the limit of vanishing active-sterile mixing it reduces to the $3\nu$ mixing matrix in the standard parameterization
(see Ref.~\cite{FNPA-ARNPS}).
\vspace{0.3cm}
\item
It keeps the first row, which gives the mixing of $\nu_{e}$, as simple as possible
and the second row, which gives the mixing of $\nu_{\mu}$, is simpler than the third.
This is useful,
because we are able to observe oscillations of electron and muon neutrinos.
\end{enumerate}
For example,
in the case of 3+1 mixing that we will consider in the following,
the $3\times4$ mixing matrix is given by
\begin{equation}
U
=
\begin{pmatrix}
c_{12}
c_{13}
c_{14}
&
s_{12}
c_{13}
c_{14}
&
c_{14}
s_{13}
e^{-i\delta_{13}}
&
s_{14}
e^{-i\delta_{14}}
\\
\cdots
&
\cdots
&
\begin{array}{c} \displaystyle
c_{13} c_{24} s_{23}
\\[-0.1cm] \displaystyle
- s_{13} s_{14} s_{24} e^{i(\delta_{14}-\delta_{13})}
\end{array}
&
c_{14}
s_{24}
\\
\cdots
&
\cdots
&
\cdots
&
c_{14}
c_{24}
s_{34}
e^{-i\delta_{34}}
\end{pmatrix}
\begin{pmatrix}
1 & 0 & 0 & 0
\\
0 & e^{i\lambda_{21}} & 0 & 0
\\
0 & 0 & e^{i\lambda_{31}} & 0
\\
0 & 0 & 0 & e^{i\lambda_{41}}
\end{pmatrix}
,
\label{U}
\end{equation}
where the dots replace the elements with long expressions.
The mixing parameters beyond the standard $3\nu$ mixing are three mixing angles
$\vartheta_{14}$,
$\vartheta_{24}$,
$\vartheta_{34}$,
two Dirac CP-violating phases
$\delta_{14}$
and
$\delta_{34}$,
and a Majorana CP-violating phases $\lambda_{41}$.
However,
the quantities measurable in experiments involving $\nu_{e}$ and $\nu_{\mu}$
depend on only two mixing angles,
$\vartheta_{14}$ and $\vartheta_{24}$,
one Dirac CP-violating phase, $\delta_{14}$,
and the Majorana CP-violating phases $\lambda_{41}$
(that however has no effect in neutrino oscillations,
as all Majorana phases).

Let us now consider neutrino oscillations.
The probability of neutrino oscillations in vacuum,
\begin{align}
P_{\nu_{\alpha}\to\nu_{\beta}}(L,E)
=
\null & \null
\delta_{\alpha\beta}
-
\underbrace{
4
\sum_{k>j}
\operatorname{Re}\!\left[
U_{{\alpha}k}^{*}
U_{{\beta}k}
U_{{\alpha}j}
U_{{\beta}j}^{*}
\right]
\sin^{2}\left( \frac{\Delta{m}^{2}_{kj} L}{4E} \right)
}_{\makebox[0pt][c]{\footnotesize
CP conserving
}}
\nonumber
\\
\null & \null
\phantom{
\delta_{\alpha\beta}
}
+
\underbrace{
2
\sum_{k>j}
\operatorname{Im}\!\left[
U_{{\alpha}k}^{*}
U_{{\beta}k}
U_{{\alpha}j}
U_{{\beta}j}^{*}
\right]
\sin\!\left( \frac{\Delta{m}^{2}_{kj} L}{2E} \right)
}_{\makebox[0pt][c]{\footnotesize
CP violating
}}
,
\label{oscprob}
\end{align}
depends on the neutrino squared mass differences
$\Delta{m}^{2}_{kj} \equiv m_{k}^2 - m_{j}^2$
and the elements of the mixing matrix $U$,
that are the fundamental physical quantities being measurement
in neutrino oscillation experiments.
Different experiments are characterized by
the source-detector distance $L$,
the neutrino energy $E$,
and the flavors $\alpha$ and $\beta$
through which a neutrino is produced and detected.
If $\beta\neq\alpha$, we have flavor transitions detected in ``appearance'' experiments.
If $\beta=\alpha$, we speak of a survival probability detected in a ``disappearance'' experiment.

Neutrino oscillations that can be explained in the standard framework of
$3\nu$ mixing\footnote{
To be more clear, it would be better to speak about a
``effective low-energy $3\nu$ mixing'',
because what we consider $3\nu$ mixing could be generated
through the seesaw mechanism
and the mixing of the active flavor neutrinos
with the heavy massive neutrinos is practically negligible.
}
have been observed with high accuracy in a variety of
solar, atmospheric and long-baseline\footnote{
By convention, long-baseline neutrino oscillation experiments are those that have a
source-detector distance and a neutrino energy band
that give sensitivity to oscillations generated by the $\Delta{m}^2_{\text{ATM}}$
in Eq.~(\ref{ATM}).
Neutrino oscillation experiments that are sensitive to larger $\Delta{m}^2$'s
are called ``short-baseline''.
}
experiments
(see the recent global fits in Refs.~\cite{deSalas:2017kay,Capozzi:2018ubv,Esteban:2018azc}).
In this framework,
there are only two independent squared-mass differences:
the solar and atmospheric $\Delta{m}^2$'s given by
\begin{align}
\null & \null
\Delta{m}^2_{\text{SOL}}
=
\Delta{m}^2_{21}
\simeq
7.4 \times 10^{-5} \, \text{eV}^2
,
\label{SOL}
\\
\null & \null
\Delta{m}^2_{\text{ATM}}
=
|\Delta{m}^2_{31}|
\simeq
|\Delta{m}^2_{32}|
\simeq
2.5 \times 10^{-3} \, \text{eV}^2
.
\label{ATM}
\end{align}
Hence,
there is a hierarchy of $\Delta{m}^2$'s,
with
$ \Delta{m}^2_{\text{ATM}} \simeq 34 \Delta{m}^2_{21} $.
By convention,
$\Delta{m}^2_{\text{SOL}}$ is assigned to $\Delta{m}^2_{21}$,
with the numbering of the neutrino mass eigenstates such that
$m_{2}>m_{1}$.
On the other hand,
$\nu_{3}$ can be either heavier than $\nu_{2}$,
in the so-called ``normal ordering''
with
$ \Delta{m}^2_{31} > \Delta{m}^2_{32} > 0 $,
or
lighter than $\nu_{1}$,
in the so-called ``inverted ordering''
with
$ \Delta{m}^2_{32} < \Delta{m}^2_{31} < 0 $
(see the review in Ref.~\cite{deSalas:2018bym}).

The short-baseline neutrino oscillation anomalies discussed in Section~\ref{sec3}
indicate that there is at least one additional squared-mass difference
\begin{equation}
\Delta{m}^2_{\text{SBL}} \sim 1 \, \text{eV}^2
,
\label{SBL}
\end{equation}
which is much larger than the solar and atmospheric
squared-mass differences.
In order to accommodate this new $\Delta{m}^2$,
the framework of neutrino mixing
must be extended with the addition of at least one light massive neutrino
in addition to the three massive neutrinos in the standard $3\nu$ mixing scheme.
In the flavor basis,
the non-standard massive neutrinos correspond to sterile neutrinos,
because the LEP measurements of the invisible width of the $Z$ boson
have shown that there are only three active neutrinos
(see Ref.~\cite{FNPA-ARNPS}).

In this review, following Okkam's razor, we consider the simplest framework
that can explain the short-baseline anomalies,
i.e. the existence of one non-standard massive neutrino
at the eV mass scale, which corresponds to a light sterile neutrino in the flavor basis.
Let us however emphasize that we consider such four-neutrino mixing scheme as an effective one,
in the sense that other non-standard massive neutrinos may exist,
corresponding to other sterile neutrinos,
but their mixing with the active neutrinos is too small to have any
observable effect in current experiments.

\begin{figure}[t]
\begin{center}
\includegraphics*[width=\textwidth]{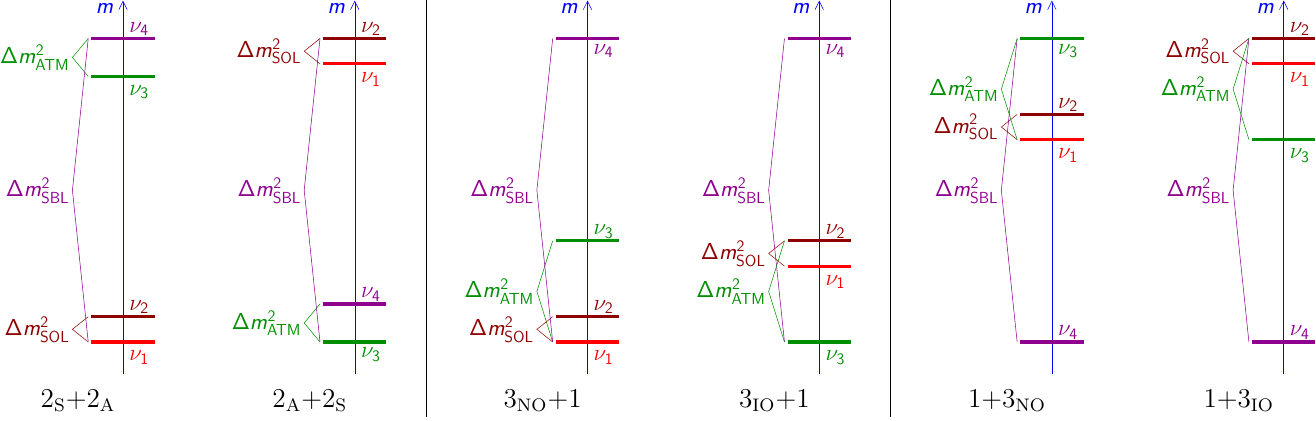}
\end{center}
\caption{\label{fig:schemes}
Schematic illustration of the
2+2, 3+1, and 1+3 neutrino mixing schemes.
}
\end{figure}

Considering only the $\Delta{m}^2$'s,
with the addition of a sterile neutrinos it is possible to have the different types of
mass spectra illustrated schematically in Fig.~\ref{fig:schemes}:

\begin{description}

\item[2+2]
In these schemes there are two pairs of massive neutrinos separated by
$\Delta{m}^2_{\text{SBL}}$.
The two pairs have mass splittings corresponding to
$\Delta{m}^2_{\text{SOL}}$
and
$\Delta{m}^2_{\text{ATM}}$,
for which there are the two possibilities shown in Fig.~\ref{fig:schemes}
($2_{\text{S}}$+$2_{\text{A}}$ and $2_{\text{A}}$+$2_{\text{S}}$).
These schemes were favored in the late 90's
(see Ref.~\cite{Bilenky:1998dt}), after the discovery of the LSND anomaly,
because they do not suffer of the so-called ``appearance-disappearance tension''
(discussed in Section~\ref{sec4} for the 3+1 and 1+3 schemes).
However,
the 2+2 schemes are strongly disfavored now by the more precise
solar and atmospheric neutrino oscillation data
\cite{Maltoni:2004ei}.
The reason is that the 2+2 schemes are not perturbations of the standard $3\nu$ mixing,
because the unitarity of the mixing matrix implies that the sterile neutrino must have large mixing
either with the $\nu_{1}$, $\nu_{2}$ pair or the $\nu_{3}$, $\nu_{4}$.
Therefore there should be large active-sterile oscillations either of solar neutrinos
or of atmospheric and long-baseline neutrinos,
but the data exclude this possibility.
For this reason, we do not discuss the 2+2 schemes any further.

\item[3+1]
In these schemes there is a new non-standard massive neutrinos $\nu_{4}$
that is heavier than the three standard massive neutrinos,
with a mass gap corresponding to
$\Delta{m}^2_{\text{SBL}}$.
These schemes are allowed by the existing solar, atmospheric and long-baseline experiments,
because they can be a perturbation of standard $3\nu$ mixing
that has small effects on the oscillations of solar, atmospheric and long-baseline neutrinos,
such that they are compatible with the existing data.
This is achieved with
\begin{equation}
|U_{\alpha 4}|^2 \ll 1
\qquad
(\alpha=e,\mu,\tau)
,
\label{smallmix}
\end{equation}
that means that the non-standard massive neutrinos $\nu_{4}$ must be mostly sterile.
In the following discussion we always assume this constraint.

\item[1+3]
In these schemes there is a new non-standard massive neutrinos $\nu_{4}$
that is lighter than the three standard massive neutrinos.
The vacuum oscillations of neutrinos in the 1+3 schemes are the same as those in the 3+1 scheme.
However,
since the mass gap between $\nu_{4}$ and $\{\nu_{1},\nu_{2},\nu_{3}\}$ corresponds to
$\Delta{m}^2_{\text{SBL}}$,
in these schemes the three standard massive neutrinos are at the eV scale.
Since the flavor neutrinos
$\{\nu_{e},\nu_{\mu},\nu_{\tau}\}$
are mainly mixed with
$\{\nu_{1},\nu_{2},\nu_{3}\}$,
the 1+3 schemes are disfavored by the cosmological upper bound on the neutrino masses,
that is smaller than 1 eV,
and by the upper bound on the effective neutrino mass in neutrinoless double-$\beta$ decay
if neutrinos are Majorana particles
(see the reviews in Refs.~\cite{Lattanzi:2017ubx,DellOro:2016tmg}).
Hence,
in the following we will not consider the 1+3 schemes.
However, one can keep in mind that all the results obtained from
experiments with neutrino oscillations in vacuum in the 3+1 schemes
apply also to the 1+3 schemes.

\end{description}

In the remainder of this review, we consider only the 3+1 schemes.
For the analysis of the short-baseline anomalies it is needed to know the
corresponding effective oscillation probabilities.
Of course, the exact oscillation probabilities are given by Eq.~(\ref{oscprob}),
but this expression is very complicated because it takes into account the effects of all the
$\Delta{m}^2$'s.
In short-baseline experiments the effects of the small
$\Delta{m}^2_{\text{SOL}}$
and
$\Delta{m}^2_{\text{ATM}}$
are negligible,
because they generate oscillations at larger distances.
Therefore, it is possible to neglect their effects and obtain the effective short-baseline
oscillation probabilities
\cite{Bilenky:1996rw}
\begin{equation}
P_{\nua{\alpha}\to\nua{\beta}}^{\text{SBL}}
=
\left|
\delta_{\alpha\beta}
-
\sin^2 2\vartheta_{\alpha\beta}
\sin^{2}\!\left( \frac{\Delta{m}^2_{41}L}{4E} \right)
\right|
,
\label{probSBL}
\end{equation}
where
$\Delta{m}^2_{41} = \Delta{m}^2_{\text{SBL}}$,
and
\begin{equation}
\sin^2 2\vartheta_{\alpha\beta}
=
4
|U_{\alpha 4}|^2
\left| \delta_{\alpha\beta} -  |U_{\beta 4}|^2 \right|
.
\label{ampSBL}
\end{equation}
These oscillation probabilities
have the same form as the oscillation probabilities
in the case of two-neutrino mixing
(see Ref.~\cite{FNPA-ARNPS})
and their amplitudes have been written in terms of the effective mixing angles
$\vartheta_{\alpha\beta}$
that depend only on the elements in the fourth column of the mixing matrix (\ref{U}),
which connect the flavor neutrinos to the non-standard massive neutrino $\nu_{4}$.
The electron and muon neutrino and antineutrino appearance and disappearance
in short-baseline experiments
depend on
$|U_{e4}|^2$ and $|U_{\mu4}|^2$,
which
determine the amplitude
\begin{equation}
\sin^22\vartheta_{e\mu}
=
4 |U_{e4}|^2 |U_{\mu4}|^2
=
\sin^22\vartheta_{14} \sin^2\vartheta_{24}
\label{sem}
\end{equation}
of
$\nua{\mu}\leftrightarrow\nua{e}$
transitions,
the amplitude
\begin{equation}
\sin^22\vartheta_{ee}
=
4 |U_{e4}|^2 \left( 1 - |U_{e4}|^2 \right)
=
\sin^22\vartheta_{14}
\label{see}
\end{equation}
of
$\nua{e}$
disappearance,
and
the amplitude
\begin{equation}
\sin^22\vartheta_{\mu\mu}
=
4 |U_{\mu4}|^2 \left( 1 - |U_{\mu4}|^2 \right)
=
\sin^22\vartheta_{24} \cos^2\vartheta_{14}
+
\sin^22\vartheta_{14} \sin^4\vartheta_{24}
\simeq
\sin^22\vartheta_{24}
\label{smm}
\end{equation}
of
$\nua{\mu}$
disappearance,
where we considered the approximation of small mixing angles given by
the constraint (\ref{smallmix}).
One must also note that the effective short-baseline oscillation probabilities
of neutrinos and antineutrinos are equal,
because they depend on the absolute values of the elements of the mixing matrix.
Therefore, even if there are Dirac CP-violating phases in the mixing matrix
[the standard $\delta_{13}$ and the non-standard $\delta_{14}$ and $\delta_{34}$
in the parameterization (\ref{U})],
CP violation cannot be measured in short-baseline experiments.
In order to measure the effects of these phases
it is necessary to perform experiments sensitive to the oscillations generated by the smaller squared-mass differences
$\Delta{m}^2_{\text{ATM}}$
\cite{deGouvea:2014aoa,Klop:2014ima,Gandhi:2015xza}
or
$\Delta{m}^2_{\text{SOL}}$
\cite{Long:2013hwa}.
For example,
considering
$|U_{e4}| \sim |U_{\mu4}| \sim |U_{e3}| \sim \varepsilon \sim 0.15$
at order $\varepsilon^3$ we have\footnote{
Assuming the very likely validity of the CPT symmetry,
the survival probabilities of neutrinos and antineutrinos are equal
(see Ref.~\cite{FNPA-ARNPS}).
Therefore,
CP violation can be observed only in oscillations between different flavors.
}
\begin{align}
P_{\nu_{\mu}\to\nu_{e}}^{\text{LBL}}
=
\null & \null
4
\,
\sin^2\vartheta_{13} \sin^2 \vartheta_{23} \sin^{2}\Delta_{31}
\nonumber
\\
\null & \null
+
2
\sin\vartheta_{13}
\sin 2\vartheta_{12}
\sin 2\vartheta_{23}
(\alpha\Delta_{31})
\sin\Delta_{31}
\cos(
\Delta_{32}
+
\delta_{13}
)
\nonumber
\\
\null & \null
+
4
\,
\sin\vartheta_{13}
\sin\vartheta_{14}
\sin\vartheta_{24}
\sin \vartheta_{23}
\sin\Delta_{31}
\sin(
\Delta_{31}
+
\delta_{13}
-
\delta_{14}
)
,
\label{probLBL}
\end{align}
where
$
\alpha
\equiv
\Delta{m}^2_{21}/|\Delta{m}^2_{31}|
\sim
\varepsilon^2
$
and
$\Delta_{kj} \equiv \Delta{m}^2_{kj} L / 4 E$.
The first term in Eq.~(\ref{probLBL}) is the dominant one, of order $\varepsilon^2$.
The other two terms are subdominant, of order $\varepsilon^3$,
but they are extremely interesting because they depend on the
standard CP-violating phase $\delta_{13}$ and,
regarding the third term,
the non-standard CP-violating phase $\delta_{14}$.
If the 3+1 mixing scheme is real, the presence of $\delta_{14}$ in Eq.~(\ref{probLBL})
can have important effects in the search of CP violation
in the current long-baseline experiments
and
in the future DUNE and Hyper-Kamiokande experiments
\cite{deGouvea:2014aoa,Klop:2014ima,Gandhi:2015xza}.

The fact that all the oscillation amplitudes in Eq.~(\ref{ampSBL})
depend only on the tree elements in the last column of the mixing matrix
implies that they are related.
In particular,
the appearance and disappearance amplitudes are related by
\cite{Okada:1996kw,Bilenky:1996rw}
\begin{equation}
\sin^2 2\vartheta_{\alpha\beta}
\simeq
\frac{1}{4}
\,
\sin^2 2\vartheta_{\alpha\alpha}
\,
\sin^2 2\vartheta_{\beta\beta}
\qquad
(\alpha,\beta=e,\mu,\tau; \alpha\neq\beta)
,
\label{appdis}
\end{equation}
in the approximation (\ref{smallmix}) of small mixing.
This relation causes the appearance-disappearance tension
of the current data discussed in Section~\ref{sec4}.
Let us emphasize that the tension cannot be alleviated
by considering more than one sterile neutrino,
because there are relations of the type (\ref{appdis})
for each additional sterile neutrino
\cite{Giunti:2015mwa}.
Physically, the tension arises because any
$\nua{\alpha}\to\nua{\beta}$
transition with $\alpha\neq\beta$ can occur only if there are corresponding
$\nua{\alpha}$ and $\nua{\beta}$ disappearances.

\section{Short-baseline neutrino oscillation anomalies}
\label{sec3}
There are three indications of neutrino oscillations in short-baseline experiments,
that are usually called ``anomalies'',
because they require the existence of a new $\Delta{m}^2$
beyond the two in standard $3\nu$ mixing,
as explained in Section~\ref{sec2}.
In the following Subsections we briefly review these three anomalies:
\ref{subs:LSND}
LSND and MiniBooNE;
\ref{subs:Gallium}
the Gallium neutrino anomaly;
\ref{subs:RAA}
the reactor antineutrino anomaly.

\subsection{LSND and MiniBooNE}
\label{subs:LSND}

The Liquid Scintillator Neutrino Detector (LSND) experiment
\cite{Aguilar:2001ty}
was performed at the Los Alamos Neutron Science Center (LANSCE)
from 1993 to 1998,
where an intense proton beam of about 1 mA and energy 798 MeV
hitting a target produced a large number of pions.
Since most of the $\pi^{-}$ were absorbed by the nuclei of the target,
the neutrinos are dominantly produced by
$\pi^{+} \to \mu^{+} + \nu_{\mu}$
and
$\mu^{+} \to e^{+} + \nu_{e} + \bar\nu_{\mu}$
decays,
most of which are decay at rest (DAR).
Since no $\bar\nu_{e}$ are produced in these two decays,
the experimental setup is ideal for searching possible
$\bar\nu_{\mu}\to\bar\nu_{e}$
oscillations.
The LSND detector was placed at a distance of 30 m from the target and consisted of a tank
filled with 167 tons of liquid scintillator viewed by photomultiplier tubes.

The energy spectrum of $\bar\nu_{\mu}$ produced by $\mu^{+}$ DAR is
$
\phi_{\bar\nu_{\mu}}(E)
\propto
E^2 \left( 3 - 4 E / m_{\mu} \right)
$
for neutrino energies $E$ smaller than
$
E_{\text{max}} = (m_{\mu}-m_{e})/2 \simeq 52.6 \, \text{MeV}
$.
The $\bar\nu_{e}$ events
have been detected through the inverse beta decay (IBD) process
\begin{equation}
\bar\nu_{e} + p \to n + e^{+}
,
\label{IBD}
\end{equation}
that has the very well known cross section (neglecting terms of order $E/M$,
where $M$ is the nucleon mass, and radiative corrections)
\cite{Vogel:1999zy,Strumia:2003zx}
\begin{equation}
\sigma_{\bar\nu_{e}p}
=
\frac{2 \pi^2}{f \tau_{n} m_{e}^5} \, E_{e} p_{e}
,
\label{IBDcs}
\end{equation}
where
$E_{e}$ and $p_{e}$ are, respectively, the positron energy and momentum,
$\tau_{n}$ is the neutron lifetime,
and
$f=1.7152$
is the phase-space factor.
Neglecting the small recoil energy of the neutron,
the neutrino energy $E$ is inferred from the measured electron
kinetic energy $T_{e} = E_{e} - m_{e}$ through the energy-conservation relation
\begin{equation}
E
=
T_{e} + m_{e} + m_{n} - m_{p}
\simeq
T_{e} + 1.8 \, \text{MeV}
,
\label{IBDene}
\end{equation}
where $m_{p}$ and $m_{n}$ are, respectively, the proton and neutron masses.

The LSND data on $\bar\nu_{\mu}\to\bar\nu_{e}$ oscillations
cover the energy range
$20 \lesssim T_{e} \lesssim 60 \, \text{MeV}$.
and show a significant excess of $\bar\nu_{e}$-like events over the background,
at the level of about $3.8\sigma$,
corresponding to an average transition probability of
$ ( 2.64 \pm 0.67 \pm 0.45 ) \times 10^{-3} $
\cite{Aguilar:2001ty}.
These oscillations can be explained with a
$\Delta{m}^2_{\text{SBL}} \gtrsim 0.1 \, \text{eV}^2$
connected with the existence of sterile neutrinos,
as explained in Section~\ref{sec2}.

The LSND anomaly has been explored in the MiniBooNE experiment
that is operating at Fermilab since 2002.
In this experiment the neutrinos are produced by the 8 GeV protons
from the Fermilab booster hitting a beryllium target
and producing a beam of pions.
The sign of the pions that are focused towards the detector is determined
by the polarity of a focusing horn.
The detector, placed at a distance of 541 m from the target,
consists of a tank filled with 818 tons of pure mineral oil (CH$_2$)
viewed by 1520 phototubes that detect the Cherenkov light and isotropic scintillation produced by charged particles.

Since in MiniBooNE the neutrino energy ranges from 200 MeV to 3 GeV
the range of $L/E$, from 0.18 to 2.7 m/MeV,
covers the LSND range of $L/E$ (from 0.5 to 1.5 m/MeV).
However, since in LSND $L/E$ is smaller than 1.5 m/MeV,
the LSND signal should be seen in MiniBooNE for $E \gtrsim 360 \, \text{MeV}$.

Initially the MiniBooNE experiment operated in ``neutrino mode''
with a focused beam of $\pi^{+}$ that decayed in a decay tunnel producing
an almost pure beam or $\nu_{\mu}$'s.
In the first article \cite{AguilarArevalo:2007it}
the MiniBooNE collaboration considered the data with $E > 475 \, \text{MeV}$,
arguing that this threshold ``greatly reduced a number of backgrounds with
little impact on the fit's sensitivity to oscillations''.
No excess over background was observed, leading to a 98\% exclusion of
neutrino oscillation as the explanation of the LSND anomaly.
However an excess of $\nu_{e}$-like events was observed below the $475 \, \text{MeV}$
analysis threshold.
This low-energy excess was confirmed in the following years,
in both neutrino \cite{AguilarArevalo:2008rc,Aguilar-Arevalo:2018gpe}
and
antineutrino \cite{Aguilar-Arevalo:2013pmq}
modes,
whereas the data above $475 \, \text{MeV}$
continued to show little or no excess over the backgrounds.
Since most of the energy range below $475 \, \text{MeV}$
correspond to values of $L/E$ outside the LSND range,
the low-energy excess is an effect different from the LSND anomaly,
and it has been considered as the
``MiniBooNE low-energy anomaly''.
A possible explanation of this anomaly is that the low-energy excess
is produced by photons,
that cannot be distinguished from $\nua{e}$-like events in the MiniBooNE detector
(single photon events are generated by
neutral-current $\nu_{\mu}$-induced $\pi^{0}$ decays in which only one of the two decay photons is visible).
This possibility is going to be investigated in the MicroBooNE experiment at Fermilab
\cite{Gollapinni:2015lca},
with a large Liquid Argon Time Projection Chamber (LArTPC)
in which electrons and photons can be distinguished.

\begin{figure}[t]
\begin{center}
\includegraphics*[width=\textwidth]{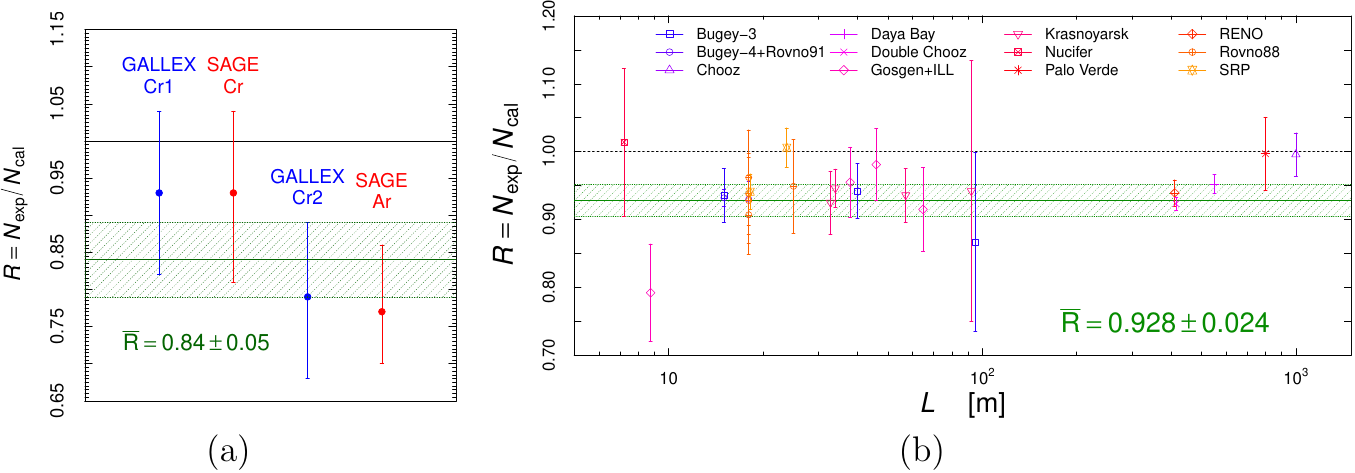}
\end{center}
\caption{\label{fig:galrea}
The Gallium neutrino (a)
and reactor antineutrino (b)
anomalies.
The data error bars represent the uncorrelated experimental uncertainties.
The horizontal solid green line and the surrounding shadowed band
show the average ratio $\overline{R}$ and its uncertainty
calculated taking into account the experimental uncertainties,
their correlations and, in panel (b), the theoretical uncertainty of the
Huber-Mueller antineutrino fluxes.
}
\end{figure}

\subsection{The Gallium neutrino anomaly}
\label{subs:Gallium}

The Gallium neutrino anomaly
\cite{Laveder:2007zz,Giunti:2006bj,Giunti:2010zu,Giunti:2012tn}
consists in the disappearance of $\nu_{e}$
measured in the
Gallium radioactive source experiments performed by the
GALLEX
\cite{Kaether:2010ag}
and
SAGE
\cite{Abdurashitov:2005tb}
collaborations.
These source experiments have been made to test the working of the
GALLEX and SAGE solar neutrino detectors.
Intense artificial ${}^{51}\text{Cr}$ and ${}^{37}\text{Ar}$ radioactive sources,
which produce $\nu_{e}$'s through the electron capture processes
$
e^{-} + {}^{51}\text{Cr} \to {}^{51}\text{V} + \nu_{e}
$
and
$
e^{-} + {}^{37}\text{Ar} \to {}^{37}\text{Cl} + \nu_{e}
$
have been placed near the center of each detector.
The $\nu_{e}$'s have been detected with the same reaction that was used for the detection
of solar electron neutrinos:
$
\nu_{e} + {}^{71}\text{Ga} \to {}^{71}\text{Ge} + e^{-}
$.

Figure~\ref{fig:galrea}a shows the
ratios $R$ of the number of
measured ($N_{\text{exp}}$)
and calculated
($N_{\text{cal}}$; see the review in Ref.~\cite{Gariazzo:2015rra})
electron neutrino events
in the two ${}^{51}\text{Cr}$ GALLEX experiments
and in the
${}^{51}\text{Cr}$ and ${}^{37}\text{Ar}$ SAGE experiments.
The average ratio shown in the figure is
$\overline{R} = 0.84 \pm 0.05$,
which shows the
$2.9\sigma$
deficit that is the Gallium anomaly.

Since the average neutrino traveling distances
in the GALLEX and SAGE
radioactive source experiments
are
$\langle L \rangle_{\text{GALLEX}} = 1.9 \, \text{m}$
and
$\langle L \rangle_{\text{SAGE}} = 0.6 \, \text{m}$,
and the neutrino energy is about 0.8 MeV,
the Gallium neutrino anomaly can be explained by neutrino oscillations
generated by a
$
\Delta{m}^2_{\text{SBL}}
\gtrsim
1 \, \text{eV}^2
$.

\subsection{The reactor antineutrino anomaly}
\label{subs:RAA}

The reactor antineutrino anomaly was discovered in 2011
\cite{Mention:2011rk}
as a consequence of the new calculation of Mueller et al. of the fluxes of $\bar\nu_{e}$'s
produced in a reactor by the decay chains of the four fissionable nuclides
$^{235}\text{U}$,
$^{238}\text{U}$,
$^{239}\text{Pu}$, and
$^{241}\text{Pu}$
\cite{Mueller:2011nm}.
The new calculation predicted fluxes that are about 5\% larger than the previous calculation
\cite{Vogel:1980bk,Schreckenbach:1985ep,Hahn:1989zr}.
The resulting expected detection rate turned out to be larger than that observed
in several short-baseline reactor neutrino experiments
with detectors placed at distances between about 10 and 100 m from the respective reactor,
generating the ``reactor antineutrino anomaly''.
In Ref.~\cite{Mueller:2011nm},
the ${}^{238}\text{U}$ antineutrino flux
was calculated with the ``ab initio'' method, using the nuclear databases,
whereas the
$^{235}\text{U}$,
$^{239}\text{Pu}$, and
$^{241}\text{Pu}$
antineutrino fluxes
have been calculated inverting the spectra of the electrons
measured at ILL in the 80's
\cite{Schreckenbach:1985ep,Hahn:1989zr,Haag:2014kia}.
This calculation was improved by Huber in Ref.~\cite{Huber:2011wv}
and the resulting fluxes including the ${}^{238}\text{U}$ of Ref.~\cite{Mueller:2011nm}
are usually called ``Huber-Mueller'' fluxes.

In reactor neutrino experiments
electron antineutrinos are
detected through the IBD process (\ref{IBD})
with the cross section (\ref{IBDcs})
in a liquid or solid scintillator detector.
The neutrino energy is obtained from the measured electron
kinetic energy trough Eq.~(\ref{IBDene}),
with a threshold of about $1.8 \, \text{MeV}$.
Since the neutrino energy spectrum decreases with energy
and the cross section (\ref{IBDcs}) increases with energy,
the energy spectrum of the detected $\bar\nu_{e}$'s
has a peak at about 3.6 MeV and decreases rapidly for larger energies,
with a tail that extends up to about 9 MeV
(see the review in Ref.~\cite{Bemporad:2001qy}).

Figure~\ref{fig:galrea}b shows the
ratios $R$ of the measured
($N_{\text{exp}}$)
and calculated
($N_{\text{cal}}$)
number of electron antineutrino events in the
Bugey-4 \cite{Declais:1994ma},
ROVNO91 \cite{Kuvshinnikov:1990ry},
Bugey-3 \cite{Declais:1995su},
Gosgen \cite{Zacek:1986cu},
ILL \cite{Kwon:1981ua,Hoummada:1995zz},
Krasnoyarsk \cite{Vidyakin:1987ue,Vidyakin:1990iz,Vidyakin:1994ut},
Rovno88 \cite{Afonin:1988gx},
SRP \cite{Greenwood:1996pb},
Chooz \cite{Apollonio:2002gd},
Palo Verde \cite{Boehm:2001ik},
Nucifer \cite{Boireau:2015dda},
Double Chooz \cite{Abe:2014bwa,Bezerra-NOW2018},
Daya Bay \cite{An:2017osx,Adey:2018qct}, and
RENO \cite{RENO:2018pwo}
experiments at the respective distance $L$ from the reactor.
As shown in the figure,
the average ratio is
$\overline{R} = 0.928 \pm 0.024$,
which indicates a
$3.0\sigma$
deficit that is the reactor antineutrino anomaly.
Given the ranges of reactor neutrino energies and source-detector distances,
the deficit can be explained by neutrino oscillations
generated by a
$
\Delta{m}^2_{\text{SBL}}
\gtrsim
0.5 \, \text{eV}^2
$.

The statistical significance of the anomaly depends on the estimated uncertainties
of the Huber-Mueller fluxes;
however, these estimated uncertainties have been put in question by the discovery of an excess
in the spectrum of detected events around 5 MeV
(often called the ``5 MeV bump'')
in the
RENO \cite{RENO:2015ksa},
Double Chooz \cite{Abe:2014bwa}
and
Daya Bay \cite{An:2015nua}
experiments.
Hence,
it is plausible that the uncertainties of the Huber-Mueller fluxes are larger than the nominal ones,
but their values are unknown
(see Refs.~\cite{Huber:2016fkt,Hayes:2016qnu}).
Therefore,
the strategy of the new reactor experiments
has been to compare the spectrum of $\bar\nu_{e}$-induced events measured at different distances
searching for the differences due to oscillations.
Interesting results have been obtained in the
NEOS \cite{Ko:2016owz}
and
DANSS \cite{Alekseev:2018efk}
experiments.

The NEOS experiment \cite{Ko:2016owz}
consisted in a 1 ton detector made of Gd-loaded liquid scintillator
located at a distance of 24 m from the $2.8 \, \text{GW}_{\text{th}}$
reactor unit 5 of the Hanbit Nuclear Power Complex in Yeonggwang, Korea (see Subsection~\ref{subs:neos} for more details).
The spectrum of $\bar\nu_{e}$-induced events
was normalized to the prediction obtained from the
Daya Bay spectrum
\cite{An:2016srz}
measured at the large distance of about 550 m,
where short-baseline oscillations are averaged out.
In this way,
the information on neutrino oscillations is independent from
the theoretical flux calculation and from the 5 MeV bump effect.
The NEOS collaboration found two almost equivalent best fits at
($\Delta{m}^{2}_{41} \simeq 1.7 \, \text{eV}^2$, $\sin^22\vartheta_{ee} \simeq 0.05$)
and
($\Delta{m}^{2}_{41} \simeq 1.3 \, \text{eV}^2$, $\sin^22\vartheta_{ee} \simeq 0.04$),
with a $\chi^2$ which is lower by 6.5 with respect to absence of oscillations.

In the DANSS experiment \cite{Alekseev:2018efk}
a highly segmented plastic scintillator detector with a volume of 1 m$^3$
is placed under an industrial $3.1 \, \text{GW}_{\text{th}}$ reactor of
the Kalinin Nuclear Power Plant in Russia
(see Subsection~\ref{subs:danss} for more details).
The detector is installed on a movable platform which allows to change the
distance between the centers of the reactor and detector from 10.7 (Up) to 12.7 m (Down).
The DANSS collaboration found that
the best fit of the Down/Up spectral ratio is obtained for
$\sin^2 2\vartheta_{ee} \simeq 0.05$
and
$\Delta{m}^2_{41} \simeq 1.4 \, \text{eV}^2$,
with a $\chi^2$ that is smaller by 13.1 with respect to the case of no oscillations.

The coincidence of the NEOS and DANSS best fits at
$\sin^2 2\vartheta_{ee} \simeq 0.04 - 0.05$
and
$\Delta{m}^2_{41} \simeq 1.3 - 1.4 \, \text{eV}^2$
is a remarkable indication in favor of short-baseline active-sterile neutrino oscillations
that updates the older reactor antineutrino anomaly
and can be considered as more robust,
since it is not based on the theoretical flux calculations.

Recently also the
PROSPECT \cite{Ashenfelter:2018iov}
and
STEREO \cite{Almazan:2018wln}
experiments
(see Subsections~\ref{subs:stereo} and \ref{subs:prospect})
have released initial data on spectral ratio measurements,
which however are still not sensitive to the NEOS/Daya Bay and DANSS best-fit region.
Another experiment,
Neutrino-4 \cite{Serebrov:2018vdw}
(see Subsection~\ref{subs:neutrino4}),
found an unexpected indication of oscillations generated by
$\Delta{m}^2_{41} \simeq 7 \, \text{eV}^2$
and
$|U_{e4}|^2 \simeq 0.1$.
This mixing is very large, in contradiction with the required inequality (\ref{smallmix}),
and
in conflict with the NEOS/Daya Bay and DANSS results,
with the exclusion limit of PROSPECT \cite{Ashenfelter:2018iov},
and with the solar neutrino upper bound for $|U_{e4}|^2$
\cite{Giunti:2009xz,Palazzo:2011rj,Palazzo:2012yf,Giunti:2012tn,Palazzo:2013me,Gariazzo:2017fdh}.
Therefore,
it is difficult to consider the Neutrino-4 as a reliable
indication in favor of short-baseline neutrino oscillations.

Interesting information on the reactor $\bar\nu_{e}$ fluxes and oscillations
came also recently from the measurement of the evolution of the event rate
with the variation of the reactor fuel composition during burnup in the
Daya Bay \cite{An:2017osx,Adey:2018qct}
and
RENO \cite{RENO:2018pwo}
experiments.
The evolution data alone disfavor neutrino oscillations as the sole explanation of the reactor antineutrino anomaly,
but the combined analysis of the Daya Bay and RENO evolution data and the absolute rates of the other experiments
in Fig.~\ref{fig:nuedis}b
leave open the possibilities of neutrino oscillations,
or a flux miscalculation,
or a combination of both
\cite{Giunti:2017yid,Giunti:2019qlt}.

\section{Global fits}
\label{sec4}
The global fits of short-baseline oscillation data addressing the anomalies discussed in
Section~\ref{sec3} in terms of active-sterile neutrino oscillations
are combined analyses of a wide variety of experimental data.
The adjective ``global'' is only indicative, because there is no established consensus on the
exact set of data that must be taken into account.
In particular, there is some variation in the global analyses performed by different groups
on the inclusion of old experimental data,
as those of old reactor experiments,
and controversial data,
as the low-energy MiniBooNE data.
Moreover,
the method of analysis of the data of old experiments (as LSND and old reactor experiments)
is not unique,
because the only available information is that in the published articles,
which is not complete.
In these cases the experimental data are analyzed with reasonable assumptions,
adjusting the relevant parameters in order to reproduce as well as possible the
results presented in the corresponding experimental publication.
Taking into account these caveats,
it is clear that the results of the global fits cannot be considered as very accurate,
but must be considered as indicative of the true solution,
whose accurate value can only be found with new experiments.

The construction of a global fit of short-baseline oscillation data
passes through the following partial stages that we discuss in the following Subsections:
\ref{subs:nue}
global analysis of SBL $\nua{e}$ disappearance;
\ref{subs:app}
global analysis of SBL $\nua{\mu}\to\nua{e}$ appearance;
\ref{subs:numu}
global analysis of SBL $\nua{\mu}$ disappearance;
\ref{subs:glo}
global appearance and disappearance analysis.

\subsection{$\nu_{e}$ and $\bar\nu_{e}$ disappearance}
\label{subs:nue}

The information on SBL $\nu_{e}$ and $\bar\nu_{e}$ disappearance is given by:

\begin{enumerate}

\item
\label{item:rea-rat}
The total event rates measured in the
Bugey-4 \cite{Declais:1994ma},
ROVNO91 \cite{Kuvshinnikov:1990ry},
Bugey-3 \cite{Declais:1995su},
Gosgen \cite{Zacek:1986cu},
ILL \cite{Kwon:1981ua,Hoummada:1995zz},
Krasnoyarsk87 \cite{Vidyakin:1987ue},
Krasnoyarsk94 \cite{Vidyakin:1990iz,Vidyakin:1994ut},
Rovno88 \cite{Afonin:1988gx},
SRP \cite{Greenwood:1996pb},
Chooz \cite{Apollonio:2002gd},
Palo Verde \cite{Boehm:2001ik},
Nucifer \cite{Boireau:2015dda},
Double Chooz \cite{Abe:2014bwa,Bezerra-NOW2018},
Daya Bay \cite{An:2017osx,Adey:2018qct}, and
RENO \cite{RENO:2018pwo}
$\bar\nu_{e}$ reactor experiments.
The deficit obtained from
the comparison of the measured event rates with the predictions based on the
theoretical calculations of the
${}^{238}\text{U}$ \cite{Mueller:2011nm,Mention:2011rk}
and
${}^{235}\text{U}$,
${}^{239}\text{Pu}$,
${}^{241}\text{Pu}$ \cite{Huber:2011wv}
reactor $\bar\nu_{e}$ fluxes
is the reactor antineutrino anomaly (RAA).
Of particular importance are the recent data of the
Daya Bay \cite{An:2017osx,Adey:2018qct}
and
RENO \cite{RENO:2018pwo}
experiments on the evolution of the event rate
with the variation of the reactor fuel composition during burnup.
These data can distinguish the neutrino oscillation solution of the RAA
from explanations based on incorrect flux predictions,
because the suppression due to neutrino oscillations is independent on the reactor fuel composition,
whereas different variations of the predictions of the four
reactor $\bar\nu_{e}$ fluxes affect the evolution of the event rate.
It turns out that the Daya Bay and RENO evolution data
disfavor the neutrino oscillation solution of the RAA
in comparison with an incorrect ${}^{235}\text{U}$ flux prediction
\cite{An:2017osx,Giunti:2017yid,RENO:2018pwo,Giunti:2019qlt}.
Hence,
taking into account also the skepticism on the
theoretical flux predictions raised from the discovery of the ``5 MeV bump''
in the spectral data of several experiments
(see Refs.~\cite{Huber:2016fkt,Hayes:2016qnu}),
it is now common to fit the reactor rates with hybrid hypotheses
involving both neutrino oscillations and corrections to the theoretical flux predictions
\cite{Giunti:2017yid,Dentler:2017tkw,Gariazzo:2018mwd,Dentler:2018sju,Giunti:2019qlt}.

\item
\label{item:rea-spe}
The ratio of the spectra measured at different distance from the reactor in the
Bugey-3 \cite{Declais:1995su} experiment,
in the
NEOS \cite{Ko:2016owz}
and
Daya Bay \cite{An:2016srz}
experiments, and in the
DANSS
experiment \cite{Alekseev:2018efk}.
These measurements provide information on neutrino oscillations that is independent from the
theoretical reactor flux predictions,
because the ratio of the spectra at different distances is independent on the
initial reactor $\bar\nu_{e}$ flux.
Hence this type of measurement is crucial for an unambiguous solution of the RAA
and is pursued in the new dedicated experiments
Neutrino-4 \cite{Serebrov:2018vdw},
PROSPECT \cite{Ashenfelter:2018iov}, and
STEREO \cite{Almazan:2018wln},
that have recently released initial data,
and the
SoLid \cite{Manzanillas:2018npw}
experiment.
The current status of the spectral ratio measurements is that
the NEOS/Daya Bay and DANSS data are in agreement on the indication of active-sterile neutrino oscillations generated by
$\Delta{m}^2_{41} \simeq 1.3 \, \text{eV}^2$
and
$|U_{e4}|^2 \simeq 0.01$
\cite{Gariazzo:2018mwd,Dentler:2018sju}.
The old Bugey-3 spectral ratio measurement
and the initial
PROSPECT and STEREO measurements exclude large mixing,
without affecting mixing as small as $|U_{e4}|^2 \simeq 0.01$.
On the other hand,
the Neutrino-4 collaboration reported an unexpected evidence of active-sterile neutrino oscillations generated by
$\Delta{m}^2_{41} \simeq 7 \, \text{eV}^2$
and
$|U_{e4}|^2 \simeq 0.1$,
which is in conflict with the NEOS/Daya Bay and DANSS results,
with the exclusion limit of PROSPECT \cite{Ashenfelter:2018iov},
and with the solar neutrino upper bound for $|U_{e4}|^2$
in the following item~\ref{item:solar}.
Therefore,
the Neutrino-4 data are typically not considered in global fits.

\item
\label{item:gallium}
The Gallium source experiment data on $\nu_{e}$ disappearance reviewed in Section~\ref{sec3}.

\item
\label{item:solar}
The solar neutrino constraint on $|U_{e4}|^2$
\cite{Giunti:2009xz,Palazzo:2011rj,Palazzo:2012yf,Giunti:2012tn,Palazzo:2013me,Gariazzo:2017fdh}.

\item
\label{item:nueC}
The ratio of the
KARMEN \cite{Armbruster:1998uk}
and
LSND \cite{Auerbach:2001hz}
$\nu_{e} + {}^{12}\text{C} \to {}^{12}\text{N}_{\text{g.s.}} + e^{-}$
scattering data at different distances from the source
\cite{Conrad:2011ce,Giunti:2011cp}.

\item
\label{item:atm}
The atmospheric neutrino constraint on $|U_{e4}|^2$
\cite{Maltoni:2007zf,Dentler:2018sju}.

\end{enumerate}

\begin{figure}[t]
\begin{center}
\includegraphics*[width=\textwidth]{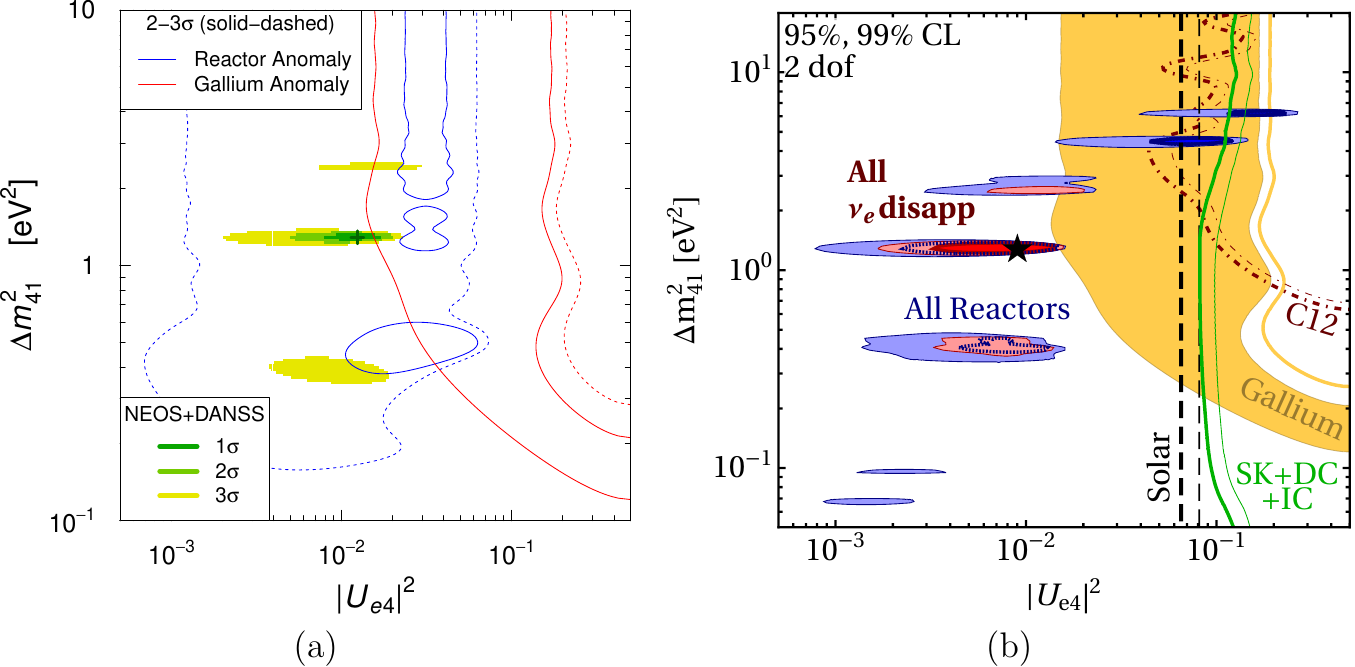}
\end{center}
\caption{\label{fig:nuedis}
Results on SBL $\protect\nua{e}$ disappearance found in Ref.~\protect\cite{Gariazzo:2018mwd} (a)
and Ref.~\protect\cite{Dentler:2018sju} (b).
The shaded regions in panel (a) have been obtained from the combined fit of
the NEOS/Daya Bay and DANSS spectral ratio data (NEOS+DANSS).
The blue and red contour lines delimit the regions allowed by the reactor and Gallium anomalies,
respectively, at $2\sigma$ (solid lines) and $3\sigma$ (dashed lines).
The blue shaded regions in panel (b) were obtained in Ref.~\cite{Dentler:2018sju}
from a global fit of the reactor neutrino data including the
NEOS/Daya Bay and DANSS spectral ratio data
and the total event rates considering as free the dominant
${}^{235}\text{U}$ and ${}^{239}\text{Pu}$
reactor $\bar\nu_{e}$ fluxes
and constraining the subdominant
${}^{238}\text{U}$ and ${}^{241}\text{Pu}$
fluxes around their theoretical predictions with a large 10\% uncertainty.
The red shaded regions have been obtained by adding the
Gallium, solar, and $\nu_{e}$-${}^{12}\text{C}$ constraints,
that are also shown separately.
The figure shows also the atmospheric neutrino constraint
obtained from the Super-Kamiokande (SK) \cite{Wendell:2010md},
DeepCore (DC) \cite{Aartsen:2017bap}
and IceCube (IC) \cite{TheIceCube:2016oqi} data,
that is comparable to the solar neutrino constraint.
}
\end{figure}

The interesting results on SBL $\nua{e}$ disappearance found in Refs.~\cite{Gariazzo:2018mwd,Dentler:2018sju}
are reproduced in Fig.~\ref{fig:nuedis}.
Figure~\ref{fig:nuedis}a shows
the allowed regions in the $|U_{e4}|^2$--$\Delta{m}^2_{41}$ plane
obtained from the combined fit of
the NEOS/Daya Bay and DANSS spectral ratio data (NEOS+DANSS).
One can see that these data are quite powerful in constraining the active-sterile
mixing parameters in a best-fit region around
$\Delta{m}^2_{41} \simeq 1.3 \, \text{eV}^2$
and
$|U_{e4}|^2 \simeq 0.01$
and constitute a model-independent indication in favor of
SBL $\nua{e}$ disappearance due to active-sterile oscillations
that is much more robust than
those of the original
reactor and Gallium anomalies,
which suffer from the dependence on the calculated reactor fluxes
and the assumed Gallium detector efficiencies.
From Fig.~\ref{fig:nuedis}a one can also see that there is a tension between
the model-independent NEOS+DANSS allowed regions
and those indicated by the reactor and Gallium anomalies
(the corresponding parameter goodness of fit
\cite{Maltoni:2003cu} are 2\% and 4\%, respectively).
This tension indicates that corrections to the theoretical reactor flux predictions are needed
and
that the efficiencies of the
GALLEX and SAGE detectors may have been overestimated.

Figure~\ref{fig:nuedis}b depicts the results obtained in Ref.~\cite{Dentler:2018sju}
from a global fit of the reactor neutrino data including the
NEOS/Daya Bay and DANSS spectral ratio data
and the total event rates considering as free the dominant
${}^{235}\text{U}$ and ${}^{239}\text{Pu}$
reactor $\bar\nu_{e}$ fluxes
and constraining the subdominant
${}^{238}\text{U}$ and ${}^{241}\text{Pu}$
fluxes around their theoretical predictions with a large 10\% uncertainty
(in order to avoid unphysical solutions).
The allowed regions in Fig.~\ref{fig:nuedis}b
are in good agreement with those depicted in Fig.~\ref{fig:nuedis}a,
because the different data sets considered in the two analyses
are dominated by the NEOS/Daya Bay and DANSS spectral ratio data
considered in both analyses.

The overall conclusions obtained from the analyses of SBL $\nua{e}$ disappearance data is that
there is an indication in favor of oscillations into sterile neutrinos
at the $3\sigma$ level, which is independent of the theoretical reactor flux calculations.
This is an improvement with respect to the original
reactor antineutrino anomaly that was based on the theoretical reactor flux calculations.
Let us however emphasize that the model-independent indication
hinges crucially on the NEOS/Daya Bay and DANSS spectral ratios
that must be confirmed by new experiments.
It is also important to emphasize that
the search for SBL $\nua{e}$ disappearance is of fundamental importance independently from the
validity or not of the indication of
$\nua{\mu}\to\nua{e}$
appearance discussed in the following subsection,
because it is possible that $|U_{e4}|^2 \simeq 0.01$,
whereas $|U_{\mu4}|^2$ is much smaller and the corresponding
$\nua{\mu}\to\nua{e}$
appearance
and
$\nua{\mu}$ disappearance
has not been seen yet.

\begin{figure}[t]
\begin{center}
\includegraphics*[width=\textwidth]{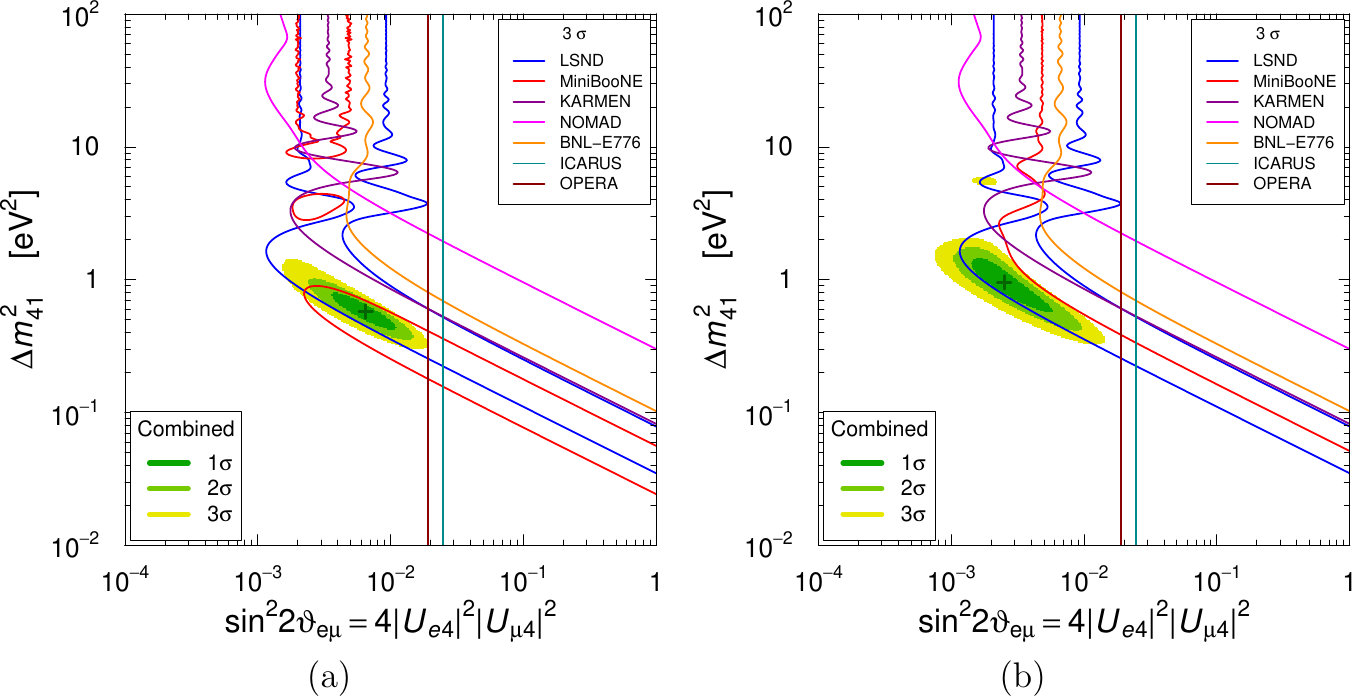}
\end{center}
\caption{\label{fig:app}
Results of SBL $\protect\nua{\mu}\to\protect\nua{e}$ appearance experiments
with all MiniBooNE data (a) and without the low-energy MiniBooNE data (b).
All the lines exclude the region on their right at $3\sigma$,
except the LSND lines in both panels and the MiniBooNE lines in panel (a) that
enclose $3\sigma$ allowed regions.
The shaded regions are allowed by the combined fit.
}
\end{figure}

\subsection{$\nu_{\mu}\to\nu_{e}$ and $\bar\nu_{\mu}\to\bar\nu_{e}$ appearance}
\label{subs:app}

Figure~\ref{fig:app} illustrates the results of all the relevant
SBL $\protect\nua{\mu}\to\protect\nua{e}$ appearance experiments:
LSND \cite{Aguilar:2001ty},
MiniBooNE \cite{Aguilar-Arevalo:2018gpe},
BNL-E776 \cite{Borodovsky:1992pn},
KARMEN \cite{Armbruster:2002mp},
NOMAD \cite{Astier:2003gs},
ICARUS \cite{Antonello:2013gut}
and
OPERA \cite{Agafonova:2013xsk}.
Of all the experiments only LSND and MiniBooNE found indications in favor of
SBL $\protect\nua{\mu}\to\protect\nua{e}$ transitions and
in Fig.~\ref{fig:app}a they have closed contours
in the plane of the oscillation parameters
$\sin^22\vartheta_{e\mu}$ and $\Delta{m}^2_{41}$.
The other experiments provide exclusion curves that constitute upper limits on
$\sin^22\vartheta_{e\mu}$
for each value of $\Delta{m}^2_{41}$.
The difference between Figs.~\ref{fig:app}a and \ref{fig:app}b is that in
Fig.~\ref{fig:app}a all the MiniBooNE data are used,
whereas Fig.~\ref{fig:app}b the controversial low-energy MiniBooNE data are omitted according to the
``pragmatic approach'' advocated in Ref.~\cite{Giunti:2013aea}.
The pragmatic approach is motivated by the fact that the low-energy MiniBooNE excess is too large to be fitted
with a small value of $\sin^22\vartheta_{e\mu}$, compatible with the bounds of other experiments,
as one can see in Fig.~\ref{fig:app}a, where the MiniBooNE contour lies at small values of
$\Delta{m}^2_{41}$
and large values of
$\sin^22\vartheta_{e\mu}$,
in tension with the ICARUS and OPERA upper bounds
and
with the disappearance bound discussed in Subsection~\ref{subs:glo}.
As discussed at the end of Subsection~\ref{subs:LSND},
most of the MiniBooNE low-energy excess
lies out of the $L/E$ range of LSND and it is probably not due to oscillations.

Comparing Figs.~\ref{fig:app}a and \ref{fig:app}b, one can see that
without the low-energy data the $3\sigma$ MiniBooNE constraint
changes from a closed contour to an exclusion curve.
As a result,
the combined allowed region without the low-energy MiniBooNE data
(Fig.~\ref{fig:app}b)
is larger than that with low-energy MiniBooNE data
(Fig.~\ref{fig:app}a)
and allows smaller values of
$\sin^22\vartheta_{e\mu}$.
This is important in the global fit appearance and disappearance data
discussed in Subsection~\ref{subs:glo},
because the disappearance data constrain severely $\sin^22\vartheta_{e\mu}$,
according to Eq.~(\ref{appdis}).

\begin{figure}[t]
\begin{center}
\includegraphics*[width=\textwidth]{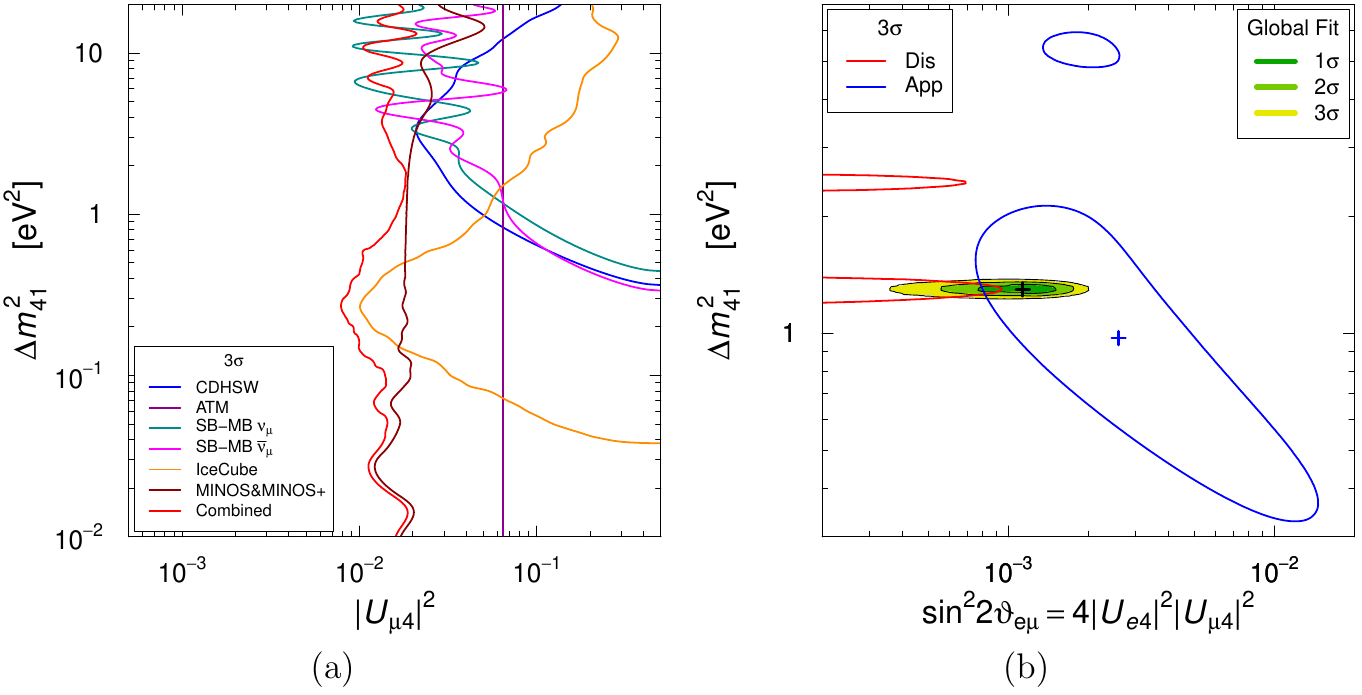}
\end{center}
\caption{\label{fig:appdis}
Results of SBL $\protect\nua{\mu}$ disappearance experiments (a)
and global fit of appearance (App) and disappearance (Dis) data (b).
All the lines in panel (a) and the Dis lines in panel (b)
exclude the region on their right at $3\sigma$.
The App lines in panel (b) enclose the $3\sigma$ allowed regions.
The shaded regions are allowed by the global combined fit.
}
\end{figure}

\subsection{$\nu_{\mu}$ and $\bar\nu_{\mu}$ disappearance}
\label{subs:numu}

If SBL active-sterile oscillations exist,
they must show up also in $\nua{\mu}$ disappearance,
as explained in Section~\ref{sec2}.
However, so far no $\nua{\mu}$ disappearance has been observed.
Figure~\ref{fig:appdis}a shows a summary of the exclusion curves found in the
$\nua{\mu}$ disappearance experiments
CDHSW \cite{Dydak:1983zq},
SciBooNE-MiniBooNE with
neutrinos \cite{Mahn:2011ea} and antineutrinos \cite{Cheng:2012yy},
IceCube \cite{TheIceCube:2016oqi},
MINOS\&MINOS+ \cite{Adamson:2017uda},
and the atmospheric neutrino bound \cite{Maltoni:2007zf}.
One can see that the recent MINOS\&MINOS+ bound is particularly severe
for $\Delta{m}^2_{41} \sim 1 \, \text{eV}^2$
and determines the overall combined limit on $|U_{\mu4}|^2$ in that region.
This strong bound causes the strong appearance-disappearance tension
discussed in Subsection~\ref{subs:glo}
\cite{Gariazzo:2018mwd,Dentler:2018sju}.

\subsection{Appearance and disappearance}
\label{subs:glo}

Figure~\ref{fig:appdis}b shows the results of the global fit of appearance and disappearance data.
The appearance data are those corresponding to Fig.~\ref{fig:app}b,
without the controversial low-energy MiniBooNE data.
In spite of this choice,
one can see that there is a strong tension between the region within the blue contours allowed
at $3\sigma$ by
the appearance data and the combined bound of
$\nua{e}$ and $\nua{\mu}$ disappearance data
that exclude at $3\sigma$ all the region outside the two red semicontours.
Although the standard goodness-of-fit is fine
($54\%$),
the appearance-disappearance parameter goodness-of-fit
is as low as
$0.015\%$,
disfavoring the global 3+1 fit at
$3.8\sigma$.
Considering a global fit with the low-energy MiniBooNE data,
we get still a favorable standard goodness-of-fit of
$21\%$,
but the appearance-disappearance parameter goodness-of-fit drops to
$2 \times 10^{-7}$,
which disfavors the global 3+1 fit at
$5.2\sigma$
(see also Ref.~\cite{Dentler:2018sju}).

Therefore, the current status of the global fit of appearance and disappearance data
indicates that the interpretation of the results of
some experiment or group of experiments in terms of neutrino oscillations is not correct.
We can envisage the following scenarios:

\begin{enumerate}

\renewcommand{\labelenumi}{(\theenumi)}
\renewcommand{\theenumi}{\Alph{enumi}}

\item
The LSND excess of $\bar\nu_{e}$-like events is not due to oscillations
and the coincidence of oscillations in NEOS/Daya Bay and DANSS is a fluke.
In this case the remaining indications in favor of neutrino oscillations
(MiniBooNE, total reactor event rates versus theoretical predictions, and the Gallium anomaly)
are rather weak and could also have other explanations, leading to the demise
of the eV-scale sterile neutrinos.

\item
The LSND excess is not due to oscillations,
but the coincidence of oscillations in NEOS/Daya Bay and DANSS is real.
In this case, the lack of observation of SBL $\nua{\mu}$ disappearance
is due to a small value of $|U_{\mu4}|^2$,
that can generate, together with $|U_{e4}|^2 \simeq 0.01$,
a SBL $\nua{\mu}\to\nua{e}$ appearance that is smaller than the LSND excess.

\item
The LSND excess is due to oscillations,
but the coincidence of oscillations in NEOS/Daya Bay and DANSS is a fluke.
In this case,
given the upper bounds on $|U_{\mu4}|^2$ found in the $\nua{\mu}$ disappearance experiments,
the LSND oscillations require a value of $|U_{e4}|^2$ that is larger than about 0.05.
This value is compatible with the Gallium anomaly and with the reactor antineutrino anomaly.

\item
The strong MINOS\&MINOS+ upper bound on $|U_{\mu4}|^2$ is inaccurate.
In this case the appearance-disappearance tension is relaxed
and there is an acceptable global fit the data.

\end{enumerate}

On a positive note,
it is likely that these scenarios are going to be explored well in the new experiments
reviewed in Section~\ref{sec6}
and we will know the truth in a few years.

\section{Other effects of light sterile neutrinos}
\label{sec5}
Light sterile neutrinos can have a wide variety of effects, besides the neutrino oscillations
that we have considered so far.
The main effects can appear in phenomena that are sensitive to the neutrino masses:
$\beta$ decay, neutrinoless double-$\beta$ decay and cosmology.
In the following subsections we briefly review the main aspect of these phenomena.

\subsection{$\beta$ decay}
\label{subs:beta}

The eV-scale mass of the non-standard massive neutrino in the 3+1 scenario
can contribute significantly to the distortion of the energy spectrum of the electron
emitted in a nuclear $\beta$ decay.
In experiments that have a sensitivity to neutrino masses of the order of 1 eV,
the Kurie function (see Ref.~\cite{FNPA-ARNPS}) can be approximated by
\begin{align}
K^2(T)
\simeq
\null & \null
\left( Q - T_{e} \right)
\sqrt{ \left( Q - T_{e} \right)^{2} - m_{\beta}^{2} }
\,
\Theta(Q-T_{e}-m_{\beta})
\nonumber
\\
\null & \null
+
\left( Q - T_{e} \right)
|U_{e4}|^2
\sqrt{ \left( Q - T_{e} \right)^{2} - m_{4}^{2} }
\,
\Theta(Q-T_{e}-m_{4})
,
\label{kurie2}
\end{align}
where $T_{e}$ is the electron kinetic energy,
$
Q
=
M_{\text{I}}
-
M_{\text{F}}
-
m_{e}
$
is the $Q$-value of the process
($M_{\text{I}}$ and $M_{\text{F}}$ are the initial and final nuclear masses;
$m_{e}$ is the electron mass),
$\Theta$ is the Heaviside step function, and
$ m_{\beta}^2 = \sum_{k=1}^{3} |U_{ek}|^2 m_{k}^2 $
is the effective light neutrino mass.
The nonstandard neutrino mass $m_{4}$ can be measured by observing
a kink of the kinetic energy spectrum of the emitted electron at
$Q-m_{4}$.
Since the currently most powerful experiments,
Mainz
\cite{Kraus:2012he}
and
Troitsk
\cite{Belesev:2013cba},
did not find such a kink for $m_{4}^2 \gtrsim 10 \, \text{eV}^2$,
there is an upper limit on $|U_{e4}|^2$
for
$ \Delta{m}^2_{41} \simeq m_{4}^2 \gtrsim 10 \, \text{eV}^2$
\cite{Giunti:2012bc}.
This limit is far from the allowed regions in Fig.~\ref{fig:nuedis},
but future experiments may reach a sensitivity to those regions.
Let us also mention that also electron capture experiments
searching for the effects of neutrino masses are sensitive to $m_{4}$
\cite{Gastaldo:2016kak}.

\begin{figure}[t]
\begin{center}
\includegraphics*[width=\textwidth]{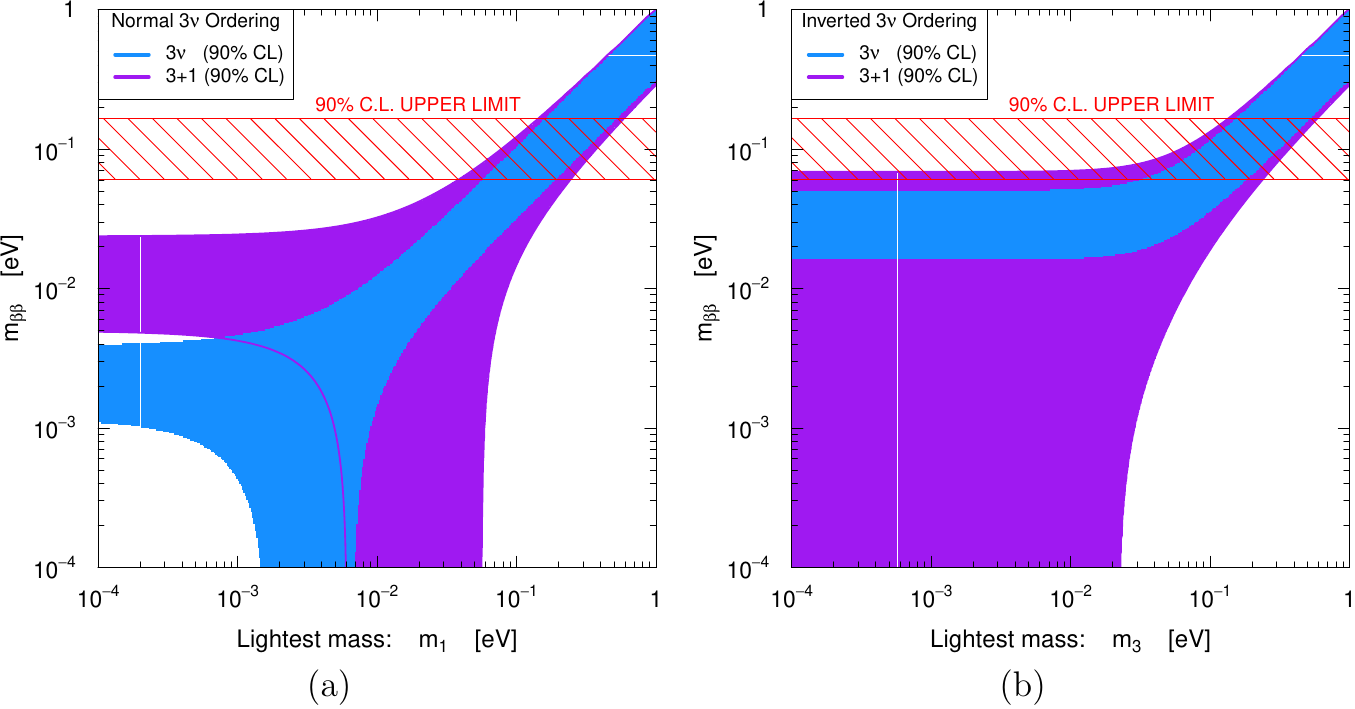}
\end{center}
\caption{\label{fig:mbb}
Value of the effective Majorana mass $m_{\beta\beta}$
as a function of the lightest neutrino mass
in the cases of $3\nu$ and 3+1 mixing with normal (a) and inverted (b) ordering
of the three lightest neutrinos.
The horizontal band is the best current
90\% C.L. upper limit for $m_{\beta\beta}$
\cite{KamLAND-Zen:2016pfg}.
}
\end{figure}

\subsection{Neutrinoless double-$\beta$ decay}
\label{subs:doublebeta}

The search for neutrinoless double-$\beta$ decay is of fundamental importance,
because it can reveal the Majorana nature of neutrinos.
The rate of neutrinoless double-$\beta$ decay
is proportional to the effective Majorana mass $m_{\beta\beta}$ given by
\begin{equation}
m_{\beta\beta}
=
\left|
|U_{e1}|^2 \, m_{1}
+
|U_{e2}|^2 \, e^{i\alpha_{2}} \, m_{2}
+
|U_{e3}|^2 \, e^{i\alpha_{3}} \, m_{3}
+
|U_{e4}|^2 \, e^{i\alpha_{4}} \, m_{4}
\right|
,
\label{mbb}
\end{equation}
where
$\alpha_{2}=2\lambda_{21}$,
$\alpha_{3}=2(\lambda_{31}-\delta_{13})$, and
$\alpha_{4}=2(\lambda_{41}-\delta_{14})$
in the parameterization of $U$ in Eq.~(\ref{U}).
The values of the $3\nu$ mixing parameters
$|U_{e1}|^2$,
$|U_{e2}|^2$, and
$|U_{e3}|^2$,
and the differences between
$m_{1}$,
$m_{2}$, and
$m_{3}$
are known from the measurements of neutrino oscillations in
solar, atmospheric and long-baseline experiments
(see the recent global fits in Refs.~\cite{deSalas:2017kay,Capozzi:2018ubv,Esteban:2018azc}).
Under the assumption of 3+1 mixing,
the value of $|U_{e4}|^2$ and the difference between
$m_{4}$ and $m_{1}$
are known from the fits of SBL $\nu_{e}$ and $\bar\nu_{e}$ disappearance data
discussed in Subsection~\ref{subs:nue}.
On the other hand,
the three phases
$\alpha_{2}$,
$\alpha_{3}$, and
$\alpha_{4}$
are totally unknown and this ignorance generates a large uncertainty
in the predicted value of $m_{\beta\beta}$.
Since $m_{4}$ is the largest mass and $|U_{e4}|^2 \simeq 0.01$ is not too small,
depending on the values of the unknown phases
the contribution of $m_{4}$ can be dominant or
there can be cancellations between the contributions of the
three standard light neutrino masses and $m_{4}$~\cite{Bilenky:2001xq,Barry:2011wb,Girardi:2013zra},
besides those intrinsic to $3\nu$ mixing
(see Ref.~\cite{DellOro:2016tmg}).
This is illustrated in Fig.~\ref{fig:mbb},
which shows the predictions for $m_{\beta\beta}$
obtained from the NEOS+DANSS allowed regions in Fig.~\ref{fig:nuedis}a,
compared with those in standard $3\nu$ mixing,
as functions of the lightest mass in the normal and inverted $3\nu$ ordering.
One can see that the contribution of $m_{4}$ can bring a dramatic change in the value of
$m_{\beta\beta}$.
In particular, in the case of normal ordering with a mass hierarchy
($m_{1} \ll \sqrt{\Delta{m}^2_{\text{SOL}}} \simeq 8 \times 10^{-3} \, \text{eV}$)
the contribution of $m_{4}$ is dominant
for any value of the unknown phases
and $m_{\beta\beta}$
is predicted to be larger than in the case of $3\nu$ mixing,
leading to the possibility to observe neutrinoless double-$\beta$ decay
with the next generation of experiments.
On the other hand,
in the case of inverted $3\nu$ ordering
the prediction for $m_{\beta\beta}$
depends crucially on the values of the unknown phases:
$m_{\beta\beta}$ can be slightly larger than in the case of $3\nu$ mixing if the phases are favorable,
but it can also be unmeasurably small if the phases generate a drastic cancellation.
Figure~\ref{fig:mbb} shows also the best current
90\% C.L. upper limit for $m_{\beta\beta}$
\cite{KamLAND-Zen:2016pfg},
with a wide band that takes into account the nuclear matrix elements uncertainties
(see Ref.~\cite{DellOro:2016tmg}).
Large values of the lightest mass are already excluded,
and the maximal 3+1 prediction in the case of inverted $3\nu$ ordering
is already partially excluded for large values of the nuclear matrix elements.

\subsection{Cosmology}
\label{subs:cosmology}

Sterile neutrinos can be produced in the early Universe
by active-sterile oscillations
before the decoupling of the active neutrinos from the primordial plasma
at a temperature of about 1 MeV,
about 1 s after the Big Bang.
However, cosmological observations exclude the full production of
a sterile neutrino in the standard $\Lambda$CDM model,
that occur if sterile neutrinos are responsible for the SBL anomalies
(see Refs.~\cite{Gariazzo:2015rra,Lattanzi:2017ubx}).
However,
the cosmological constraint can be evaded by suppressing active-sterile oscillations
in the early Universe with non standard effects,
such as ``secret'' interactions of sterile neutrinos~\cite{Hannestad:2013ana,Dasgupta:2013zpn}.
Hence,
the cosmological information on sterile neutrinos is intriguing and inspiring,
but it cannot be considered as robust as the information obtained in laboratory experiments,
which are repeatable under controlled circumstances.
The quest for the existence of sterile neutrinos must be pursued
with laboratory experiments and a positive result will require a modification of the cosmological model.

\section{Future perspectives}
\label{sec6}
Worldwide experimental efforts towards the search for light sterile neutrinos have been constantly growing since 2011. Preliminary results excluding part of the allowed range of parameters as well as fiercely debated hints for new oscillation signals have already come out. A selection of current and future perspectives for verifying the short-baseline neutrino oscillation anomalies (see Section~\ref{sec3}) and perhaps revealing sterile neutrinos is described in this section. 

\subsection{NEOS}
\label{subs:neos}
NEOS~\cite{Ko:2016owz} is a $\sim0.8$ ton Gd-based liquid scintillator experiment located in the tendon gallery of reactor unit 5 of the 2.8~GW$_{th}$ reactor unit of the Hanbit Nuclear Power Complex in Yeong-gwang, Korea. The detector is located at $23.7$ m from the center of the reactor core, and covered by an overburden of about 20 meters of water equivalent. Being a commercial nuclear station, the reactor core is not compact, leading to an average of the oscillation signal. Fast neutron background rejection is ensured using  pulse shape discrimination (PSD) on the delayed signal. Thanks to the great reactor power, the IBD counting rate is 1976 events/day leading to a high S/B ratio of 22. The NEOS detector energy resolution of $5\%$ at 1~MeV, allowing for a suitable spectral analysis.
In order to reduce the reactor neutrino fluxes uncertainties the analysis is performed by comparing the recorded spectrum with the one measured by the Daya Bay collaboration \cite{An:2016srz}.
Because Daya Bay being is on a different site,
this procedure may lack of robustness and lead to uncontrolled systematic uncertainties. NEOS took data from  August 2015 to May 2016 and reported sizable limits on the sterile neutrino parameters. In addition they reported a hint for an oscillation pattern in the ratio of the NEOS and Day Bay spectra (see Subsections~\ref{subs:RAA} and \ref{subs:nue}), but this shall be considered with a grain of salt. The NEOS detector resumed data taking since September 2018 \cite{NEOS-ICHEP2018}.

\subsection{STEREO}
\label{subs:stereo}
STEREO~\cite{Almazan:2018wln} is a Gd-based liquid scintillator running experiment located at the 58~MW High Flux Reactor of the Institute Laue-Langevin (France). The reactor core is compact and highly enriched in $^{235}$U. The 2~m$^3$ detector is installed from 9.4~m to 11.1~m away from the compact core. It consists of six optically separated cells, readout  by 4 photo-multiplier tubes from the top, surrounded by passive and active shielding. STEREO makes use of PSD technique on the delayed signal to suppress backgrounds induced by fast neutrons and achieve a S/B ratio of 0.9. To be independent from the reactor neutrino spectrum uncertainties, STEREO analyses the ratios of prompt signal spectra of different cells to that of the first cell.
STEREO is a running experiment that already released first results based on 66 days of reactor turned on and 138 days of reactor turned off. Based on this data-set, STEREO excluded a significant portion of the sterile neutrino parameter space.

\subsection{PROSPECT}
\label{subs:prospect}
PROSPECT~\cite{Ashenfelter:2018iov} is a $^6$Li-based liquid scintillator (LS) running experiment (0.1\% doping in mass) installed at the High Flux Isotope Reactor (HFIR) at Oak Ridge National Laboratory (USA), a  high power of the compact reactor (85~MW) providing mostly electron anti-neutrinos from the fission of $^{235}$U. Located at 6.7~m from the the reactor compact core,
the detector is a 2.0~m$\times$1.6~m$\times$1.2~m rectangular volume containing $\sim4$ tons of LS, allowing for a strong pulse shape discrimination (PSD) on both the prompt and delayed inverse-beta-decay signals. Thin reflecting panels divide the LS volume into an 11$\times$14 two-dimensional array of 154 optically isolated rectangular segments read out by two photo-multiplier tubes each. The energy resolution is 4.5~\% at 1~MeV.
PROSPECT successfully took the challenge of installing the detector at the Earth's surface. Thanks to its PSD and 3D reconstruction of events the collaboration managed a clear separation of signal events, with a S/B ratio of 1.36, collecting 771 IBD events per day.
In order to be independent from the reactor neutrino spectrum uncertainties, PROSPECT uses ratios of the measured IBD spectra at different baselines, normalized to the baseline-integrated measured spectrum. 
Prospect is an on-going experiment that already reported measured ratios consistent with the no oscillation hypothesis and exclude a significant portion of the sterile neutrino parameters (including the best-fit value of the reactor neutrino anomaly, for instance). 

\subsection{Neutrino-4}
\label{subs:neutrino4}

Neutrino-4~\cite{Serebrov:2018vdw}  is a $1.8\,\text{m}^3$ Gd-based liquid scintillator experiment consisting of 50 liquid scintillator sections (ten rows with 5 sections in each), installed near the powerful (100~MW) and compact SM-3 research reactor at Dimitrovgrad (Russia). The detector is installed on a movable platform, and the baseline ranges from 6 to 12 meters. The position of the detector with respect to the reactor core is changed frequently, allowing a partial cancellation of systematic uncertainties. Being located very close to the surface, the cosmogenic induced backgrounds are significant, and could not be mitigated via pulse-shape discrimination. Therefore the S/B ratio is only 0.54. Moreover, the energy resolution is only 16\% at 1~MeV. In order to be independent from the reactor neutrino spectrum uncertainties the analysis is performed by comparing the spectra recorded at the various distances of each section to a spectrum averaged over all detector sections, assuming equal efficiencies. 
As the main result, the obtained $L/E$ dependence of the IBD rate normalized to the rate averaged over all distances fits well with an oscillation signal with the following parameters: $\sin^22\vartheta_{ee}=0.35$ and $\Delta m_{41}^2=7.22~{\rm eV}^2$. The collaboration quotes a $3\sigma$ significance for the best fit point. It is worth noting that those findings are already in tensions with the limits obtained by the other reactor measurements, like PROSPECT \cite{Ashenfelter:2018iov}, and with the solar neutrino upper bound for $\sin^22\vartheta_{ee}$
\cite{Gariazzo:2017fdh}.

\subsection{DANSS}
\label{subs:danss}

DANSS~\cite{Alekseev:2018efk,Danilov:2018kjo} is a 1~m$^3$ plastic scintillator running experiment. The fiducial volume of the detector is highly segmented, being composed of 2500 scintillator strips coated with a $0.2$~mm thin Gadolinium surface to enhance neutron capture signals. The detector is surrounded by both active and passive shieldings and located below the $3.1 \, \text{GW}_{\text{th}}$ reactor core of the Kalinin Nuclear Power Plant (KNPP) in Russia. It is therefore protected from cosmic rays by a depth of about 50 meters of water equivalent, providing a strong suppression of the cosmogenic backgrounds (a factor of 6 in the muon flux) and their variations according to the evolution of the atmospheric pressure. The whole DANSS setup is installed on a movable platform switching regularly between three positions three times a week, providing baselines of 10.7~m, 11.7~m, and 12.7~m, respectively. Thanks to the high reactor power and to the reduced baseline, the DANSS statistics is extremely high, the detector counting 4910 events per day at the 10.7~m position from the reactor. Therefore DANSS achieves a very high signal/background ratio of more than 33 at the top location. As drawbacks DANSS's energy resolution of $\sigma_E/E \sim 34\%$ at $E=1$~MeV as well as the large extension of the core (3.7 m in height and 3.2 m in diameter) lead to some sizable smearing of the oscillation pattern, that is partially compensated with the large statistics, however.
To be independent from the reactor neutrino spectra predictions, the analysis is done by comparing the shape of the recorded energy spectra at 10.7~m and 12.7~m baselines. This strategy allows also to reduce the impact of the detector modeling uncertainties. DANSS already excluded an area of mixing parameters covering large fractions of the regions indicated by the Gallium and reactor anomalies. As a benchmark for comparison with other experiments the most preferred value $\Delta m_{41}^2=2.3~\rm{eV}^2,~\sin^22\vartheta_{ee} =0.14$~\cite{Mention:2011rk} is excluded at more than 5$\sigma$ CL \cite{Alekseev:2018efk,Danilov:2018kjo}.

\subsection{Solid}
\label{subs:solid}

SoLid~\cite{Abreu:2018ajc,Manzanillas:2018npw} is an experiment in preparation using 12800 cells made of cubes of polyvinyltoluene (PVT) of (5$\times$5$\times$5)~cm$^3$ in dimension, partially coated with thin sheets of $^6$LiF:ZnS(Ag) to capture and detect neutrons. The detector is installed at the surface level,
about 6~m away from the 60~MW SCK$\cdot$CEN BR2 research reactor in Belgium. A 288 kg prototype detector was deployed in 2015 and collected data during the operational period of the reactor and during reactor shut-down. The detector energy resolution is modest, $\sigma_E/E \sim 14\%$ at $E=1$~MeV. Moreover, because of the high number of cells, the electronics, data acquisition, and the calibration are real challenges. This extremely high segmentation allows 3D reconstruction and background-tagging, however. The full SoLid detector was commissioned at the beginning of 2018 at the BR2 nuclear plant. With an expected 41\% efficiency and 2 years running from early 2016 (300 days/year) a total of 250k events can be collected, which is sufficient to cover the current reactor anomaly region below 5 eV$^2$ at better than 99\% CL.
The detector is now operational and taking data in stable conditions~\cite{Abreu:2018ekw,Manzanillas:2018npw}.

\subsection{Best}
\label{subs:best}
BEST~\cite{Gavrin:2017iix} is a source-based experiment in preparation, aiming to search for an electron neutrino disappearance signal with a 3 MCi artificial source of electron neutrinos from $^{51}$Cr. The experiment is using the Gallium-Germanium neutrino telescope facility at the Baksan Neutrino Observatory of the INR RAS (GGNT), which has been used since 1990 for solar neutrino measurements in the SAGE experiment~\cite{Abdurashitov:2005tb}. The detector target containing 50 tons of liquid metal gallium is divided into two zones. Neutrinos are detected through the neutrino capture reaction on $^{71}$Ga and the number of interactions in the two zones is determined by counting the number of produced $^{71}$Ge atoms. BEST has the great potential to search for transitions of active neutrinos to sterile states with $\Delta m_{41}^2 \sim 1~\rm{eV}^2$ with a unique method that is mostly insensitive to the usual radioactive and cosmogenic background sources.

\subsection{KATRIN}
\label{subs:katrin}
The Karlsruhe Tritium Neutrino (KATRIN) experiment~\cite{Arenz:2018kma} is a large-scale effort to probe the absolute neutrino mass scale with a sensitivity of 0.2 eV (90\% C.L.), via a precise measurement of the endpoint spectrum of tritium $\beta$-decay. The first physics run will be performed in 2019. Eventually the study of the shape of the $\beta$-spectrum down to 100 eV below the endpoint allows for a search of light sterile neutrinos. Preliminary studies indicate that the KATRIN sensitivity can exceed that of the reactor experiments for $\Delta m_{41}^2 \gtrsim 2~\rm{eV}^2$
\cite{Kleesiek:2018mel,LasserreNeutrino2018}.
Also heavier sterile neutrinos will be searched by KATRIN in a dedicated run, TRISTAN, that will occur after the neutrino mass measurement from 2023~\cite{Mertens:2018vuu}.

\subsection{$^{163}$Ho experiments}
\label{subs:holmium}
$^{163}$Ho electron capture experiments~\cite{Gastaldo:2016kak} primarily performed to measure the neutrino mass are also sensitive to electron neutrino mixing with a sterile neutrino. In particular ECHO-1M, that is expected to collect up to 10$^{17}$ events, will allow to explore part of the 3+1 mixing parameter space indicated by the global analysis of short-baseline neutrino oscillation experiments. 

\subsection{The Fermilab SBN program}
\label{subs:SBNfermilab}

The Short-Baseline Neutrino (SBN) program~\cite{Tufanli:2017mwt} is a set of neutrino  experiments at the Fermilab laboratory, situated in the Booster Neutrino Beamline. The main goal is the search for 
$\nua{\mu}\to\nua{e}$
oscillations
and $\nua{\mu}$ disappearance
at $\Delta m_{41}^2 \sim 1~\rm{eV}^2$. The setup consists of three liquid argon time projection chamber (LAr-TPC) detectors, the Short-Baseline Near Detector (SBND), the MicroBooster Neutrino Experiment (MicroBooNE), and the Imaging Cosmic And Rare Underground Signals (ICARUS). The first physics goal is to perform a definitive test for the LSND and MiniBooNE sterile neutrino oscillation anomalies. By studying the baseline dependence of the appearance data, the SBN program will cover 99\% of the LSND-allowed region with more than 5$\sigma$ significance. In addition, the SBN experiment will extend the search for muon neutrino disappearance by an order of magnitude with respect to the existing results. 

\subsection{JSNS$^2$}
\label{subs:JSNS2}

The JSNS$^2$ (J-PARC Sterile Neutrino Search at J-PARC Spallation Neutron Source) experiment~\cite{Ajimura:2017fld} will search for light sterile neutrinos with $\Delta m_{41}^2 \sim 1~\rm{eV}^2$ at the J-PARC Materials and Life Science Experimental Facility. An intense neutrino beam is produced from muon decay at rest, after the interaction of 3 GeV protons incident (1 MW beam) on a spallation neutron target. The detector is composed of 17 tons of Gd-doped liquid scintillator and is located 24 meters away from the mercury target. The JSNS$^2$ experiment is focusing on the anti-electron neutrino appearance channel, via the detection of electron anti-neutrinos through IBD. This project will be a direct and ultimate test of the LSND experiment~\cite{Aguilar:2001ty}.

\subsection{IceCube}
\label{subs:icecube}

Light sterile neutrinos mixing with the active neutrino states can lead to deviations in the atmospheric neutrino flux with respect to that in the standard $3\nu$ mixing. The IceCube Neutrino Observatory~\cite{Terliuk:2017jyw}, located at the  South Pole, is an instrumented cubic kilometer Cherenkov neutrino detector measuring the atmospheric neutrinos with an energy threshold of 10 GeV and measuring energies up to 100 TeV.  One the one hand, IceCube uses events with energies above 400 GeV and searches for a resonant enhancement of the sterile neutrino mixing for muon antineutrinos crossing the Earth's core \cite{TheIceCube:2016oqi}. On the other hand, IceCube uses lower energy data of DeepCore \cite{Aartsen:2017bap}, a denser part of IceCube, in the energy range between 6 and 56 GeV to search for deviations from the standard atmospheric neutrino oscillations due to the sterile neutrino mixing \cite{PhysRevD.85.093010}.
Using one year of the full 86-string detector configuration, IceCube provides strict constraints on the allowed sterile neutrino mixing with muon and tau neutrinos in a 3+1 model
\cite{TheIceCube:2016oqi}. In the near future the full data available in IceCube could lead to unprecedented constraints on sterile neutrino mixings. 

\section{Conclusions}
\label{sec7}
The existence of sterile neutrinos
is one of the most intriguing possibilities
in the search for new physics beyond the Standard Model.
Exciting indications have come from some experiments
searching for short-baseline neutrino oscillations
that obtained ``anomalous'' results that cannot
be explained in the standard framework of $3\nu$ mixing,
pointing to extended frameworks with sterile neutrinos.
Following Okkam's razor, it is wise to consider the simplest 3+1 extension,
with a non-standard massive and mostly sterile neutrino at the eV mass scale.
In this approach we do not exclude the existence of other
non-standard massive and mostly sterile neutrinos
(for example the very heavy neutrinos in the seesaw mechanism).
We only assume that their mixing with the active neutrinos is so small that their effects
are negligible in current neutrino oscillation experiments.
Hence, the 3+1 scheme can be considered as an effective framework for the study
of short-baseline neutrino oscillations.

The discovery in 2011 of the reactor antineutrino anomaly,
after the older LSND and Gallium anomalies,
triggered the start of an intense experimental activity aimed at
understanding the origin of these anomalies.
Some of the new experiments have already obtained interesting results,
restricting the parameter space of active-sterile neutrino mixing.
Among them most notable are the model-independent positive
NEOS/Daya Bay and DANSS indications of short-baseline $\bar\nu_{e}$ disappearance
and the powerful
MINOS\&MINOS+ constraints on short-baseline $\nua{\mu}$ disappearance.
As a result,
the global fit of short-baseline data
has a severe appearance-disappearance tension
that needs to be resolved by the new experiments which already started data taking or are under preparation.

Several new reactor neutrino experiments will explore in the next years
short-baseline $\bar\nu_{e}$ disappearance
with precision and accuracy,
making measurements of the energy spectrum at different distances
in order to obtain information on neutrino oscillations
that are independent of the neutrino flux calculations.
Since the original reactor antineutrino anomaly may be due,
at least partially, to miscalculations of the reactor neutrino fluxes,
the main objective for the new reactor experiments is the check of the
model-independent NEOS/Daya Bay and DANSS indication
in favor of $\bar\nu_{e}$ disappearance.
Assuming the very likely validity of the CPT symmetry,
the disappearances of
$\nu_{e}$ and $\bar\nu_{e}$ are equal,
and a reactor $\bar\nu_{e}$ disappearance
must be observed also in source experiments with $\nu_{e}$'s.

The LSND $\bar\nu_{\mu}\to\bar\nu_{e}$ appearance signal
is going to be checked in the Short Baseline Neutrino experiment at Fermilab,
but it is important to keep in mind that
$\nua{\mu}\to\nua{e}$
oscillations are not possible without the corresponding
$\nua{e}$ and $\nua{\mu}$ disappearances.
Therefore,
if the LSND result is confirmed,
its explanation with neutrino oscillations requires
a relaxation of the current appearance-disappearance tension.
This can be obtained in the following two ways:
\begin{enumerate*} [1) ]%
\item
confirming the MINOS\&MINOS+ upper bound on $|U_{\mu4}|^2$
(either with a negative result or
with a positive observation of $\nua{\mu}$ disappearance below the MINOS\&MINOS+ bound)
and
finding
$\nua{e}$ disappearance with a value of $|U_{e4}|^2$ that is larger than the 0.01 indicated by
NEOS/Daya Bay and DANSS,
at least about 0.05;
\item
confirming the NEOS/Daya Bay and DANSS indication of
$\Delta{m}^2_{41} \simeq 1.3 \, \text{eV}^2$
and
$|U_{e4}|^2 \simeq 0.01$
and
finding $\nua{\mu}$ disappearance for $|U_{\mu4}|^2 \simeq 0.05$,
that is larger than the
MINOS\&MINOS+ upper bound.
\end{enumerate*}

We look forward for the exciting experimental program of the next years
that promises to unravel in a definitive way the puzzle of the
short-baseline neutrino oscillation anomalies
and enlighten us on the existence of eV-scale sterile neutrinos.

\section*{DISCLOSURE STATEMENT}
The authors are not aware of any affiliations, memberships, funding, or financial holdings that
might be perceived as affecting the objectivity of this review.

\section*{ACKNOWLEDGMENTS}
C.G. would like to thank S. Gariazzo, M. Laveder, Y.F. Li, B.R. Littlejohn and P.T. Surukuchi for collaboration and many useful discussions.



\end{document}